\documentclass[10pt,journal,compsoc]{IEEEtran}
\ifCLASSOPTIONcompsoc
  \usepackage[nocompress]{cite}
\else
  \usepackage{cite}
\fi

\usepackage{graphicx}
\usepackage{amsmath}
\usepackage[T1]{fontenc}

\usepackage{amssymb}
\usepackage{subcaption}
\usepackage{enumitem}
\usepackage{animate}
\usepackage{comment}
\usepackage[numbers, sort&compress]{natbib}
\usepackage{url}
\usepackage{xcolor}
\usepackage{capt-of}
\usepackage{etoolbox}
\usepackage[utf8]{inputenc} 
\usepackage{url}            
\usepackage{booktabs}       
\usepackage{amsfonts}       
\usepackage{nicefrac}       
\usepackage{microtype}      
\usepackage{xcolor}         
\usepackage{multicol}
\usepackage{diagbox}

\usepackage[pagebackref,breaklinks,colorlinks]{hyperref}

\usepackage[capitalize]{cleveref}
\crefname{section}{Sec.}{Secs.}
\Crefname{section}{Section}{Sections}
\Crefname{table}{Table}{Tables}
\crefname{table}{Tab.}{Tabs.}

\setlist{leftmargin=*}

\usepackage[ruled]{algorithm2e} 

\SetAlFnt{\small}
\SetAlCapFnt{\small}
\SetAlCapNameFnt{\small}
\SetAlCapHSkip{0pt}

\newcommand{\bfn}{{\bf n}}

\newcommand{\bfz}{{\bf z}}

\newcommand{\randall}[1]{\textcolor{pink}{\textbf{Randall: #1}}}

\begin{document}
\title{DeepTensor: Low-Rank Tensor Decomposition \\ with Deep Network Priors}

\author{Vishwanath~Saragadam,
        Randall~Balestriero,~\IEEEmembership{Member, IEEE,}
        Ashok~Veeraraghavan,~\IEEEmembership{Fellow,~IEEE,}
        and~Richard~G.~Baraniuk,~\IEEEmembership{Fellow,~IEEE}
\IEEEcompsocitemizethanks{\IEEEcompsocthanksitem V. Saragadam, R. Balestriero, A. Veeraraghavan, and R. Baraniuk are with the Department
of Electrical and Computer Engineering, Rice University, Houston,
TX, 77005.\protect\\
E-mail: vishwanath.saragadam@rice.edu}
\thanks{}}

%
%

\markboth{DeepTensor}%
{Shell \MakeLowercase{\textit{et al.}}: Bare Demo of IEEEtran.cls for Computer Society Journals}
%



\IEEEtitleabstractindextext{%
\begin{abstract}
DeepTensor is a computationally efficient framework for low-rank decomposition of matrices and tensors using deep generative networks.
We decompose a tensor as the product of low-rank tensor factors (e.g., a matrix as the outer product of two vectors), where each low-rank tensor is generated by a deep network (DN) that is trained in a \emph{self-supervised} manner to minimize the mean-square approximation error.
Our key observation is that the implicit regularization inherent in DNs enables them to capture nonlinear signal structures (e.g., manifolds) that are out of the reach of classical linear methods like the singular value decomposition (SVD) and principal components analysis (PCA).
Furthermore, in contrast to the SVD and PCA, whose performance deteriorates when the tensor’s entries deviate from additive white Gaussian noise, we demonstrate that the performance of DeepTensor is robust to a wide range of distributions.
We validate that DeepTensor is a robust and computationally efficient drop-in replacement for the SVD, PCA, nonnegative matrix factorization (NMF), and similar decompositions by exploring a range of real-world applications, including hyperspectral image denoising, 3D MRI tomography, and image classification.
In particular, DeepTensor offers a 6dB signal-to-noise ratio improvement over standard denoising methods for signal corrupted by Poisson noise and learns to decompose 3D tensors 60 times faster than a single DN equipped with 3D convolutions.

\end{abstract}

\begin{IEEEkeywords}
Tensor Decomposition, Matrix Factorization, Low-Rank Completion, Deep Network, Self-Supervised Learning.
\end{IEEEkeywords}}

\maketitle

\IEEEdisplaynontitleabstractindextext

%
\IEEEpeerreviewmaketitle

\IEEEraisesectionheading{\section{Introduction}\label{sec:introduction}}

\IEEEPARstart{L}{ow-rank} representations of matrices and tensors are truly ubiquitous and applied across all fields of science and engineering, from statistics~\cite{schonemann1966generalized,wold1987principal,jain2013low,candes2008exact,drineas2004clustering} to control systems~\cite{zecevic2005global,benner2013low} to computer vision~\cite{liang2012repairing,ji2010robust,zhang2013hyperspectral}, and beyond.
Low-rank representations seek to represent a large matrix/tensor as a product of smaller (and hence lower rank) matrices/fibers.
%
%
For instance, the classical approach to representing matrices in a low-rank manner is via the {\em singular value decomposition} (SVD), which expresses a matrix as a product of two smaller orthonormal matrices (containing the singular vectors) and a diagonal matrix (containing the singular values).
%
%
Thresholding the singular values creates a matrix that inhabits a lower dimensional subspace.
The SVD is a pervasive technique for data preprocessing and dimensionality reduction across a wide entire range of machine learning (ML) applications, including {\em principal component analysis} (PCA) and data whitening.
%

%
%

No matter how powerful or pervasive, however, the SVD and PCA are not without their shortcomings.
SVD/PCA is an optimal low-rank decomposition technique only under a narrow set of assumptions on the statistics of the signal and noise in the task at hand~\cite{eckart1936approximation}.
When the signal or noise is non-Gaussian, the resulting decomposition is {\em not} optimal and results in a subspace that differs from the true low-rank approximation of the underlying matrix.
These issues have been addressed somewhat successfully in the past with several signal- and application-specific regularizers that include sparsity on error~\cite{xu2010robust,goldfarb2014robust,waters2011sparcs}, total variation penalty~\cite{ji2016tensor,he2015total,wang2017hyperspectral}, and data-driven approaches~\cite{zhao2020deep,ke2020deep}.
The key observation is that a good signal model can act as a strong regularizer for estimating the low-rank factors.
Unfortunately, finding a useful signal model/regularizer for a new application can be a daunting task.
%
Does there exist a generalized regularizer that can encompass a large class of signals and applications?
We found the answer to be hidden implicitly in deep networks (DNs). 


\begin{figure*}[!tt]
    \centering
    \begin{subfigure}[b]{0.35\textwidth}
        \centering
        \includegraphics[width=\textwidth]{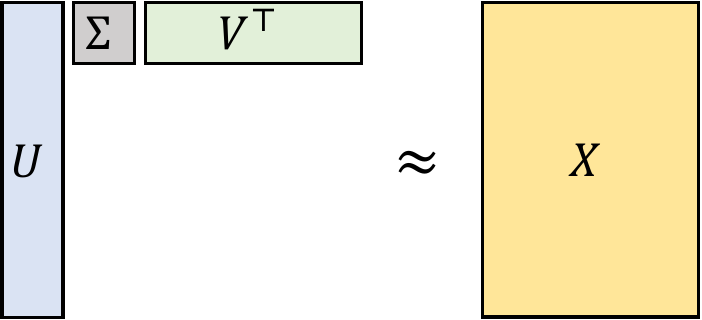}
        \caption{Low-rank representation via SVD}
    \end{subfigure}
    \hspace{2em}
    \begin{subfigure}[b]{0.55\textwidth}
        \centering
        \includegraphics[width=\textwidth]{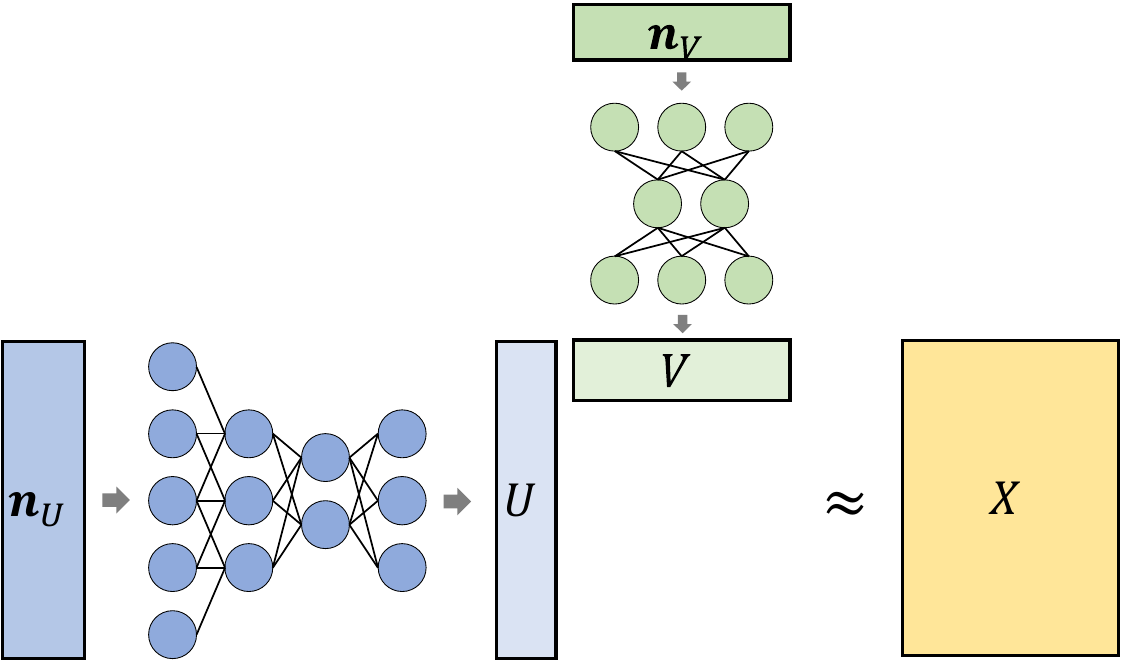}
        \caption{Low-rank representation via DeepTensor}
    \end{subfigure}
    \caption{\textbf{DeepTensor} is a new a low-rank decomposition technique that exploits the implicit regularization capabilities of DNs. Conventional low-rank factorization such as (a) SVD relies on linear factors to represent the matrix. DeepTensor represents via factor matrices that are outputs of DNs. Factorization is then achieved by learning parameters of the network in a self-supervised manner that reduces the ($\ell_2$) loss between the factor and input matrix.}
    \label{fig:overview}
\end{figure*}

\begin{figure}[!tt]
	\centering
	\includegraphics[width=\columnwidth]{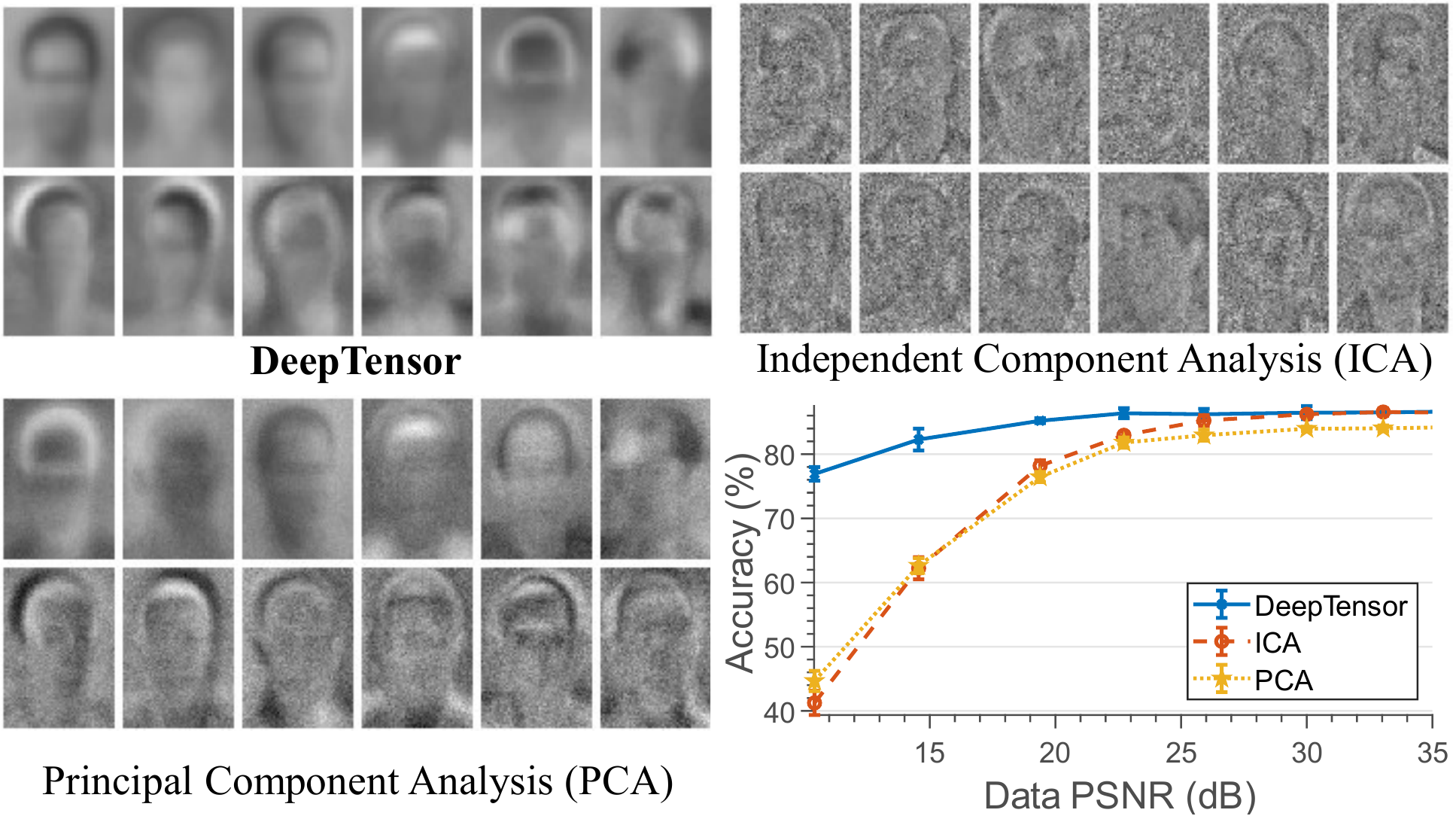}
	\caption{\textbf{DeepTensor eigenfaces.} DeepTensor is a drop-in replacement for most dimensionality reduction techniques and is robust to a wide range of signal and noise models. Here we show an ``eigenface'' decomposition with facial data corrupted by Poisson noise. The top twelve bases vectors visualized above at 15 dB input PSNR show that DeepTensor learns principal components that are significantly lower in noise levels compared to ICA or PCA. This in turn affects the classification accuracy (bottom right) that involves taking projection of input data on the principal components followed by a linear SVM classifier.}
	\label{fig:pca}
\end{figure}

{\em In this paper, we propose \textbf{DeepTensor}, a new approach for low-rank matrix/tensor decomposition that is robust to a wide class of signal and noise models.}
Our core enabling observation is that 
DNs produce signals that are implicitly regularized due to the networks' inherent inductive bias.
%
We exploit DNs as priors by representing a matrix/tensor in terms of factors output from a set of {\em untrained} generative networks (see Fig.~\ref{fig:overview}).
The parameters of the networks are learned in a self-supervised manner for each matrix/tensor using a simple MSE loss between the matrix/tensor and the product of the deeply generated factors.
%
%
The inductive bias of the generative networks enables DeepTensor to better identify the underlying low-dimensional subspace while rejecting noise, resulting in a more accurate estimate of the noise-free matrix/tensor.
	
%
%
%

{\em DeepTensor is a computationally efficient, drop-in replacement for many existing matrix/tensor factorization approaches that combines the simplicity of low-rank decomposition with the power of deep generative networks.}
Further, DeepTensor can significantly improve the performance of downstream tasks, such as image classification, that rely on low dimensional representation.
We empirically back these claims via experiments on a wide variety of real-world tasks, including denoising with low-rank approximation, ``eigenfaces"~\cite{turk1991eigenfaces} for facial recognition (see Fig.~\ref{fig:pca}), solving linear inverse problems with compressively sensed videos, tensor denoising, and recovery of 3D volumes from computed tomographic (CT) measurements.
We also highlight DeepTensor's computational efficiency and scalability for large and higher-order tensor decomposition by demonstrating that it offers a $60\times$ or more speedup for decomposing 3D tensors as compared to a single generative network equipped with full 3D convolutions.

\section{Background and Prior Work}\label{section:prior}
DeepTensor leverages classical work on low-rank approximation and recent self-supervised learning techniques with DNs~\cite{ulyanov2018deep,heckel2018deep}. We provide a brief overview of these topics to introduce our notation and intuition and to set context for our work.

{\bf Low-rank approximation.}~
Low-rank approximation seeks to represent a matrix $X\in\mathbb{R}^{M\times N}$ of rank $R=\min(M,N)$ as a product of two smaller matrices, $U\in\mathbb{R}^{M\times k}, V\in\mathbb{R}^{N \times k}$ where $k$ is generally taken to be smaller than $R$.
The specific constraints on $U, V$, and the desired objective give rise to different types of low-rank approximation algorithms.
%
%
For example, one recovers PCA  \cite{pearson1901liii}, nonnegative matrix factorization (NMF) \cite{lawton1971self} and $k$-means \cite{macqueen1967some} via
\begin{align}
&\min_{U,V}\|X-UV^T\|_F \text{~~s.t.~~ $U=V^T$}&&\hspace{-1cm}\text{(PCA/SVD)}\label{eq:svd0}\\
&\min_{U,V}\|X-UV^T\|_F \text{~~s.t.~~ $U\geq 0,V\geq 0$}&&\text{(NMF)}\\
&\min_{U,V}\|X-UV^T\|_F \text{~~s.t.~~ $[V]_{.,k}\in\{e_1,\dots,e_n\}$}&&\text{($k$-means)}\label{eq:constraints}
\end{align}
where $e_k$ is the $k^{\rm th}$ Euclidean canonical basis vector, and $[V]_{.,k}$ is the $k^{\rm th}$ column of $V$.
Applications of low-rank approximation are extremely diverse ranging from denoising \cite{tufts1993estimation,zhuang2016fast}, compression \cite{yuan2005projective}, clustering for anomaly detection \cite{xu2015anomaly}, and forecasting \cite{barratt2021low}.

The Achilles' heel of low-rank approximation is non-Gaussian signal and/or noise statistics.
There are been many extensions and variants of the SVD  algorithms, such as Robust PCA \cite{xu2010robust} that improves robustness in learning the low-rank matrices against outliers in the data.
%
Similarly, other non-Gaussian noise settings, different metrics and constraints could be employed to obtain the most adapted $U,V$ decomposition to solve the task at hand. This can be well understood based on the generative models corresponding to low-rank approximation techniques such as Probabilistic PCA \cite{tipping1999probabilistic} from which it is clear that PCA is optimal under a Gaussian noise model, and in the presence of say a Laplacian noise, an $\ell_1$ reconstruction loss should be used instead.
%
Prior work identified approaches to tackling tensor decomposition with unknown noise statistics via Bayesian optimization~\cite{zhao2015bayesian,zhao2015bayesian2}, modeling noise distribution as a mixture of Gaussian~\cite{chen2016robust}, and using a decomposition inspired by Kronecker product~\cite{bahri2018robust,xie2017kronecker}.
While the approaches are promising for several tensor decomposition applications, none of the previous works leverage inductive bias of DNs.

There has been some research in exploiting trained DNs for matrix factorization especially in the settings of nonnegative matrix factorization~\cite{pmlr-v32-trigeorgis14}, magnetic resonance imaging (MRI) denoising~\cite{ke2020deep}, and tensor completion~\cite{zhao2020deep}.
The key idea is that the statistics of training data can help regularize the inverse problem of matrix factorization.
Such techniques however suffer from dataset biases, and require very large pools of data to be effective.	

{\bf Deep networks as implicit regularizers.}~
DNs have emerged very rapidly from classification and regression applications where they have reached superhuman performance across a wide range of datasets and tasks \cite{lecun2015deep}. More recently, the use of DNs has diversified -- an illustrative and important example for this paper is the Deep Image Prior (DIP) model \cite{ulyanov2018deep}. In this setting, a DN $f$ is used as a constrained projection of a random noise vector $z$ to fit a target sample $x$ as follows
\begin{align}
\min_{\Theta}\|f_{\Theta}(z)-x \|_2^2.
\label{eq:DIP}    
\end{align}
When the architecture of the DN is carefully picked, the estimation of the input $x$ is denoised, and well reconstructed even in the presence of missing values. From the implicit regularization field it is understood that the above problem is equivalent to some problem
\begin{align}
\min_{W}\|Wz-x \|_2^2+R(W),    \label{eq:regularization}
\end{align}
where $R(W)$ is a regularization term that directly depends on the choice of the DN architecture
\cite{gunasekar2018implicit}. The key result that we will leverage throughout this paper is that {\em searching for 
the DN parameters producing the desired result in (\ref{eq:DIP}) is equivalent to searching in the space of regularizers in (\ref{eq:regularization})}.

\textbf{Related work combining DNs and low-rank decomposition.}
Applying DNs in a self-supervised manner for matrix decomposition has received surprisingly little attention.
The closest work is by
\citet{aittala2019computational}, which regularized a matrix factorization for a specific video imaging problem using generative networks.
The video sequence was represented by a generative network equipped with 3D convolutional kernels, while the light transport matrix was represented as a linear combination of the input video sequence multiplied by another generative network equipped by 2D convolutional kernels.
%
%
%
DeepTensor is in many ways inspired by the factorization idea proposed by \citet{aittala2019computational} but goes beyond light transport matrices and can be applied to a wide variety of problem settings.
%

A related but different approach to DeepTensor is the work by \citet{bacca2021compressive}, who identified that hyperspectral images (which are modeled as 3D tensors) can be represented as the output of a single generative network equipped with 3D convolutional kernels.
%
The work by \citet{bacca2021compressive} is a promising framework for solving inverse problems in hyperspectral imaging, but are not aimed at low-rank matrix factorization -- which is the focus of this paper.
The key difference between their work and DeepTensor is that the input to their 3D network is a low-rank Tucker tensor; in contrast we \emph{output} a low-rank Tucker tensor.

\section{DeepTensor Decomposition}\label{section:dlrp}
%

\subsection{Low-rank decomposition with deep network prior}
\label{sec:formulation}

\begin{figure}[!tt]
	\centering
	\begin{subfigure}{0.7\linewidth}
	\includegraphics[width=\columnwidth]{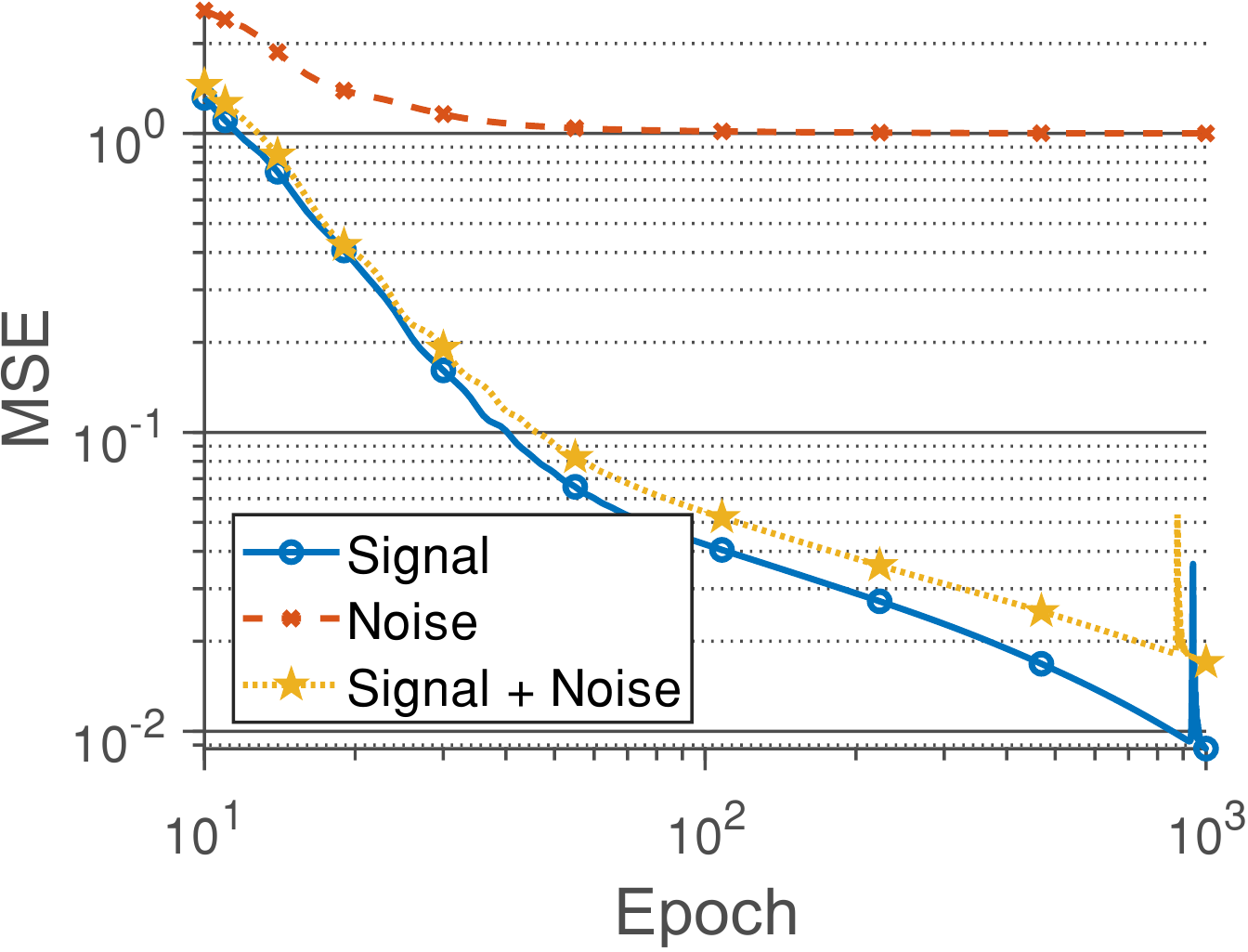}
	\end{subfigure}
	\caption{\textbf{Implicit bias of DeepTensor rejects noise.} The inductive bias due to the network architecture fits better to signal and rejects noise. To test this, we performed a rank-20 decomposition of a hyperspectral image~\cite{arad_and_ben_shahar_2016_ECCV}, iid Gaussian noise with a standard deviation of 0.01 units, and a noisy signal obtained by adding the hyperspectral image and the Gaussian noise. We observe that through the training epochs, the signal is better fit by the network, while the noise has a slower fit.}
	\label{fig:fit_dip_20}
\end{figure}

{\bf Matrix decomposition.}
Consider the low-rank decomposition in \eqref{eq:svd0}. If we include regularizers for $U$ and $V$, we obtain the following optimization function
\begin{align}
    \min_{U, v} \| X - UV^\top \|^2 + R(U, V),\label{eq:svd}
\end{align}
where $R(U, V)$ is a regularizer for $U, V$. Instead of having explicit regularizers on the left and right matrices, we model $U, V$ as the outputs of generative networks $f_U, f_V$, which yields
\begin{align}
    \min_{\theta_U, \theta_V} \| X - f_U(z_U) f_V(z_V)^\top \|^2,
\end{align}
where $z_U$ and $z_V$ are randomly initialized inputs to the networks $f_U, f_V$ respectively that have parameters $\theta_U, \theta_V$. The networks'output are of the same shape as the desired $U$ and $V$ matrices from (\ref{eq:svd}).
We note here that there is no further regularizer on the matrices -- any regularization comes from the implicit prior of the DN itself, which makes it an appealing choice to solve a diverse type of signals and noise settings. Figure~\ref{fig:fit_dip_20} visualizes the regularization offered by DNs for the task of rank-20 decomposition. Over 100 iterations, the error for signal reduces by two orders of magnitude, while the error reduces by less than one order for noise.
This slow fitting to noise is a result of implicit bias of DNs which is leveraged by DeepTensor.
%

%
%
{\bf Tensor decomposition.}
The task of tensor decomposition finds numerous applications and is an active area of research, where the major difficulty rises from defining an appropriate regularizer/basis constraint (recall (\ref{eq:constraints})).
Any constraint on the factor matrices can be expressed as regularization functions. Hence, given the following general decomposition problem
$$
\min_{U,V,\dots,W}\|X-\underbrace{U\otimes V \otimes  \dots \otimes W}_{k \text{ times}} \|^2+R(U,V,\dots,W),
$$
with $X$ a $k$-dimensional tensor, one needs to specify the correct regularizer ($R$) such that the produced decomposition is adapted for the task at hand. This search is mostly understood in narrowly defined settings such as Gaussian noise and Gaussian latent space factors. Instead, DeepTensor seeks to solve the the following optimization to compute the decomposition
\begin{align}
\min_{\theta_U,\theta_V,\dots,\theta_W}\|X-\underbrace{f_{U}(z_U)\otimes f_{V}(z_V) \otimes  \dots \otimes f_{W}(z_W) }_{k \text{times}}\|^2,\label{eq:dlrp}
\end{align}
where $f_{U},f_{V},\dots,f_{W}$ are $k$ DNs that are fed with input vectors $z_U, z_V,\times,z_W$ with parameters $\theta_U,\theta_V,\theta_W$ respectively.
Note that this decomposition closely resembles the parallel factor analysis (PARAFAC); one can instead formulate a Tucker representation which includes an extra core tensor. 
%


\begin{figure}[t!]
	\centering
	\begin{subfigure}{0.7\linewidth}
	\includegraphics[width=\linewidth]{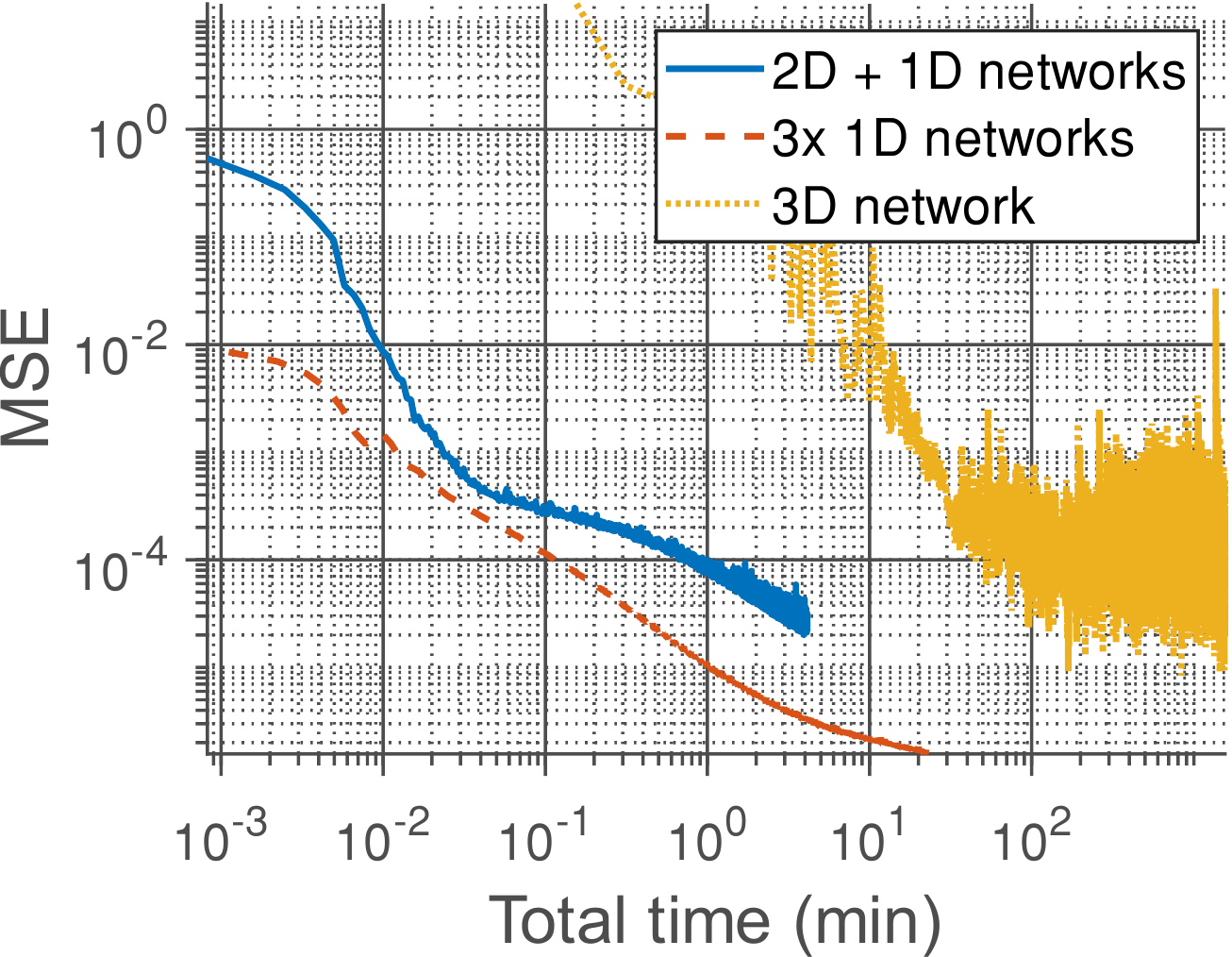}
	\end{subfigure}
	\caption{\textbf{Separable convolutions leads to speedup.} An added benefit of deep low rank decomposition is that the convolutions are separated. Hence for a 3D volume, we can exploit one 2D and one 1D network for factorization instead of a 3D network -- resulting in two orders of magnitude speed up.
	}
	\label{fig:timing}
\end{figure}

{\bf Dimensionality splitting brings tractability without sacrificing performance.}
%
While it is possible to represent the matrix itself via a DN prior, doing so will require a 2D network instead of two 1D networks. 
%
%
The goal of this section is to study the exponential computational time gain offered by separating the dimensionality of the tensor and doing the tensor products of multiple independent DN outputs as per (\ref{eq:dlrp}) versus doing the tensor product in the latent space and then using as done for the $2$-dimensional case in \cite{ulyanov2018deep} and that naturally generalizes to
\begin{equation}
f_{U,V,\dots,W}(z_U\otimes z_V\otimes  \dots \otimes z_W) .\label{eq:alternative}
\end{equation}
To understand the benefits of those two different cases we approximate a 3D magnetic resonance imaging (MRI) volume of size $128\times128\times150$ with three types of networks --- one 2D and 1D that is a natural parametrization that captures the dependence of the first two dimensions $ f_{U,V}(z_{U,V})\otimes f_{W}(z_W)$, 
three 1D (\ref{eq:dlrp}), and one 3D networks (\ref{eq:alternative}).
Figure~\ref{fig:timing} shows the plot of mean squared error (MSE) as a function of time for the three approaches.
We note that using three 1D networks is faster than using a 2D and 1D network, and both are two orders of magnitude faster than using one 3D network. Additionally, our (separate) parametrization has the benefit of keeping interpretability since one has access to each generated low-rank matrix that combine to form the observed data matrix $X$.
The choice between using networks equipped with 2D and 1D convolutions, and networks equipped with 1D convolutions is specific to the task at hand.
Tensors such as videos and hyperspectral images benefit from networks equipped with 2D and 1D convolutions.
In contrast, tensors from multi-dimensional face databases~\cite{vasilescu2002multilinear} or computed tomographic (CT) images benefit from a full tensor decomposition.
%

\begin{figure}[!tt]
	\centering
	\includegraphics[width=0.8\columnwidth]{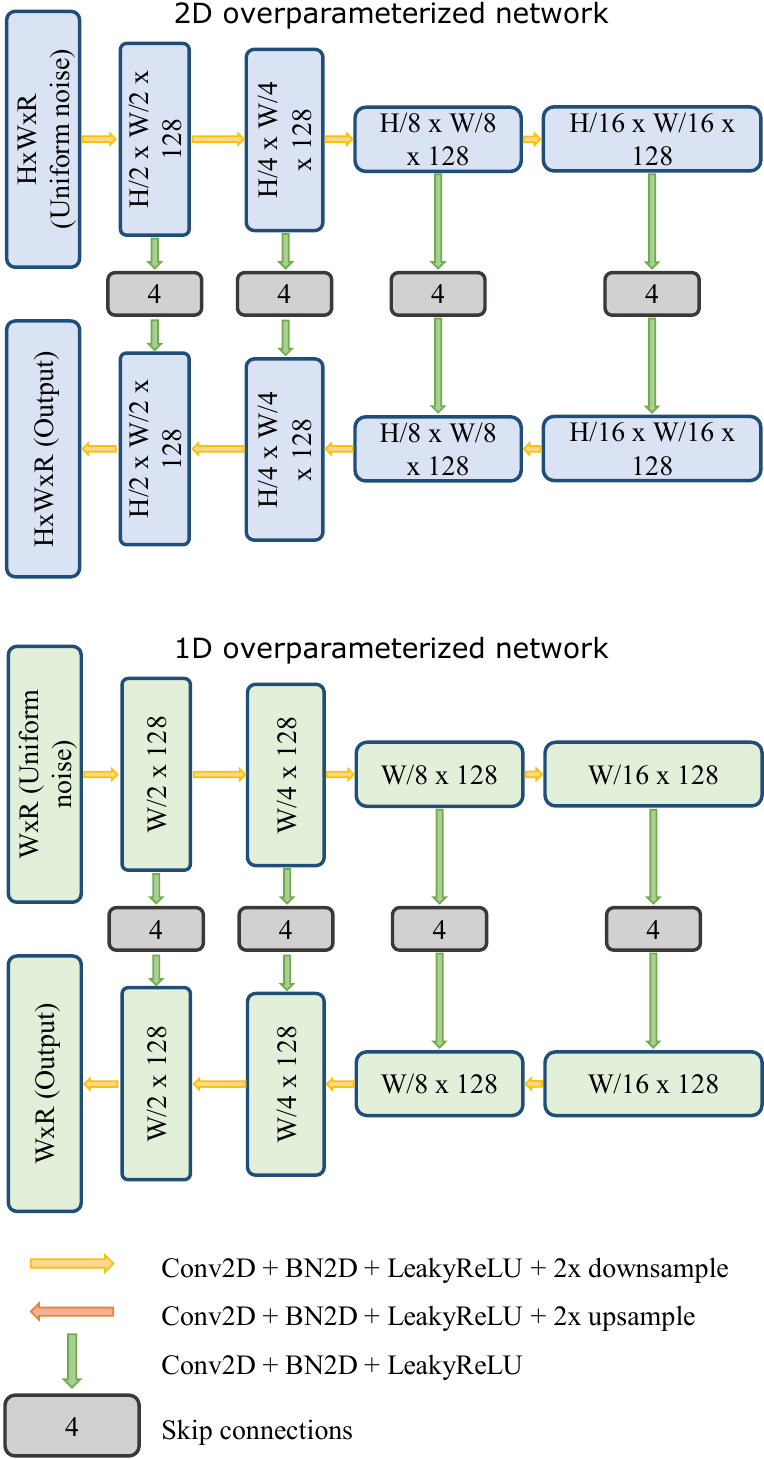}
	\caption{\textbf{Architectures for various experiments in the paper.} We used modified 2 dimensional (top) and 1 dimensional (bottom) versions of the networks proposed in Deep Image Prior~\cite{ulyanov2018deep}.}
	\label{fig:architectures}
\end{figure}

Unless explicitly specificed, we utilize overparameterized networks for the factor matrices, similar to DIP~\cite{ulyanov2018deep} (see Fig.~\ref{fig:architectures}).
However, it is possible to use underparameterized networks instead, similar to the deep decoder architecture~\cite{heckel2018deep}.
The advantage of the latter is that the learned parameters can be used as the compressed version of the tensor being decomposed. 
In contrast, for pure data imputation and denoising the DIP version should be preferred.
We compare the two choices in the upcoming section.
%
%
%
%
%
%

\begin{figure*}[!tt]
	\centering
	\begin{subfigure}[t]{0.31\textwidth}
		\centering
		\caption*{Gaussian matrix with Gaussian noise}
		\includegraphics[width=\columnwidth]{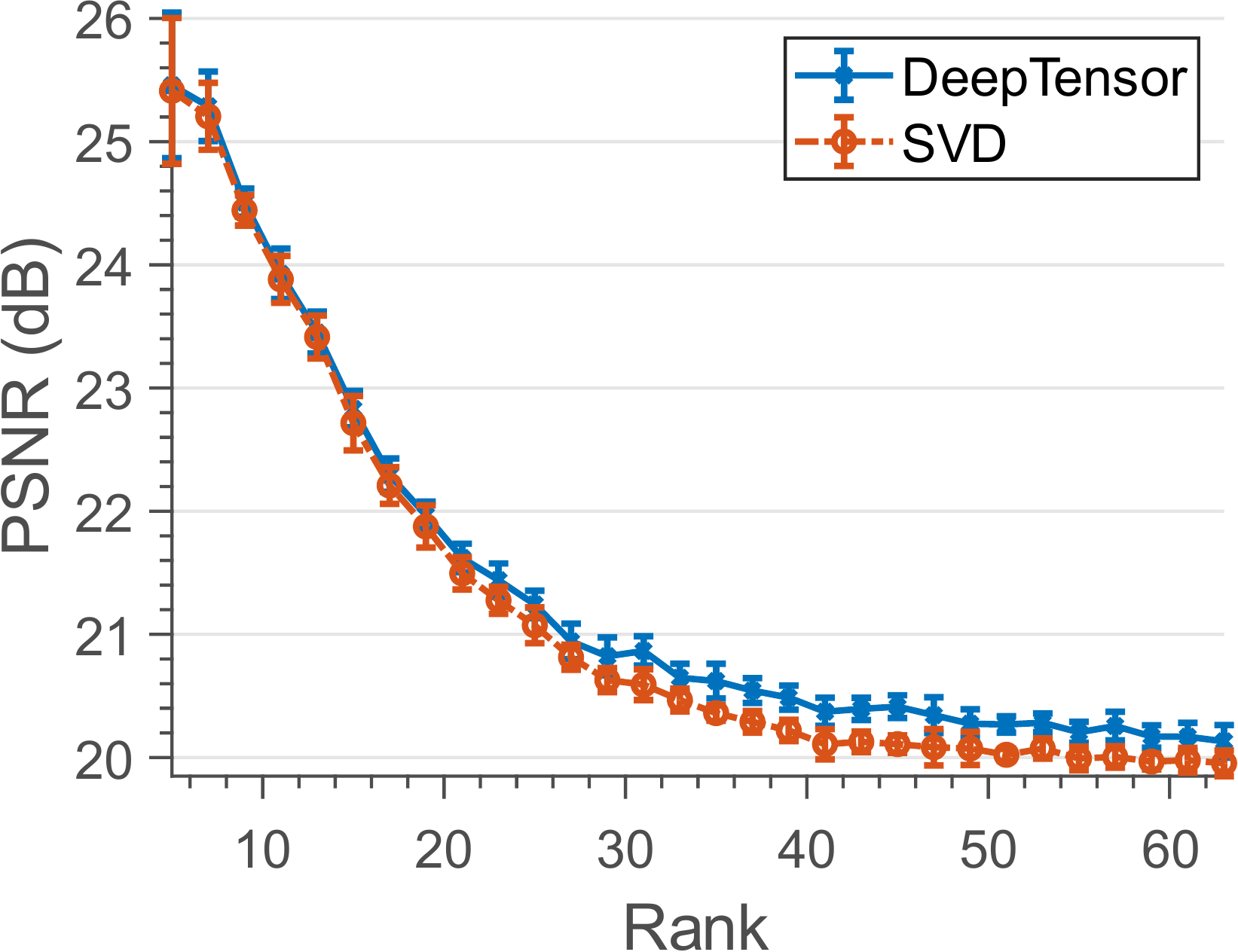}
	\end{subfigure}
	\hspace{0.5em}
	\begin{subfigure}[t]{0.31\textwidth}
		\centering
		\caption*{Piecewise constant matrix with Poisson noise}
		\includegraphics[width=\columnwidth]{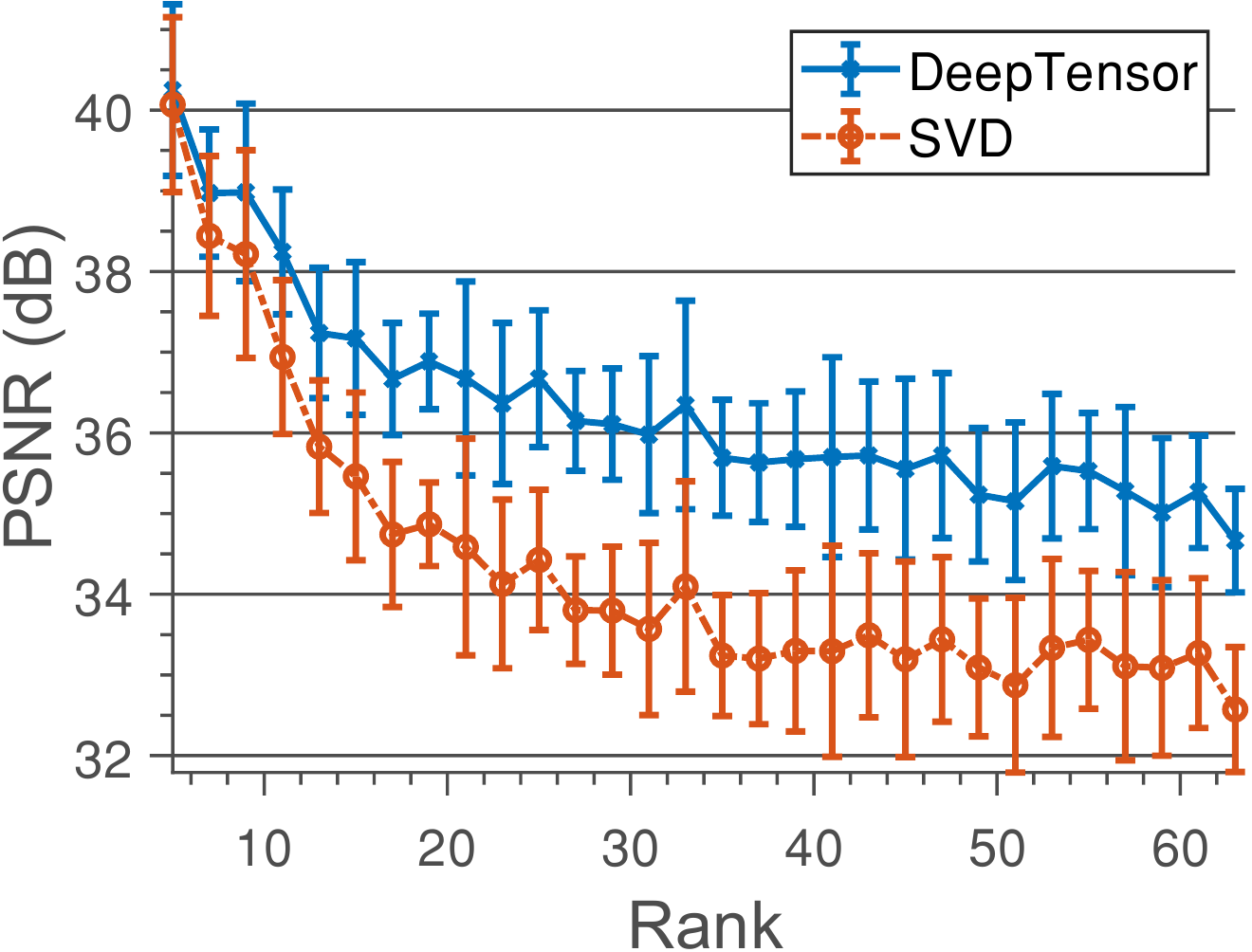}
	\end{subfigure}
	\hspace{0.5em}
	\begin{subfigure}[t]{0.31\textwidth}
		\centering
		\caption*{Piecewise constant matrix with Rician noise}
		\includegraphics[width=\columnwidth]{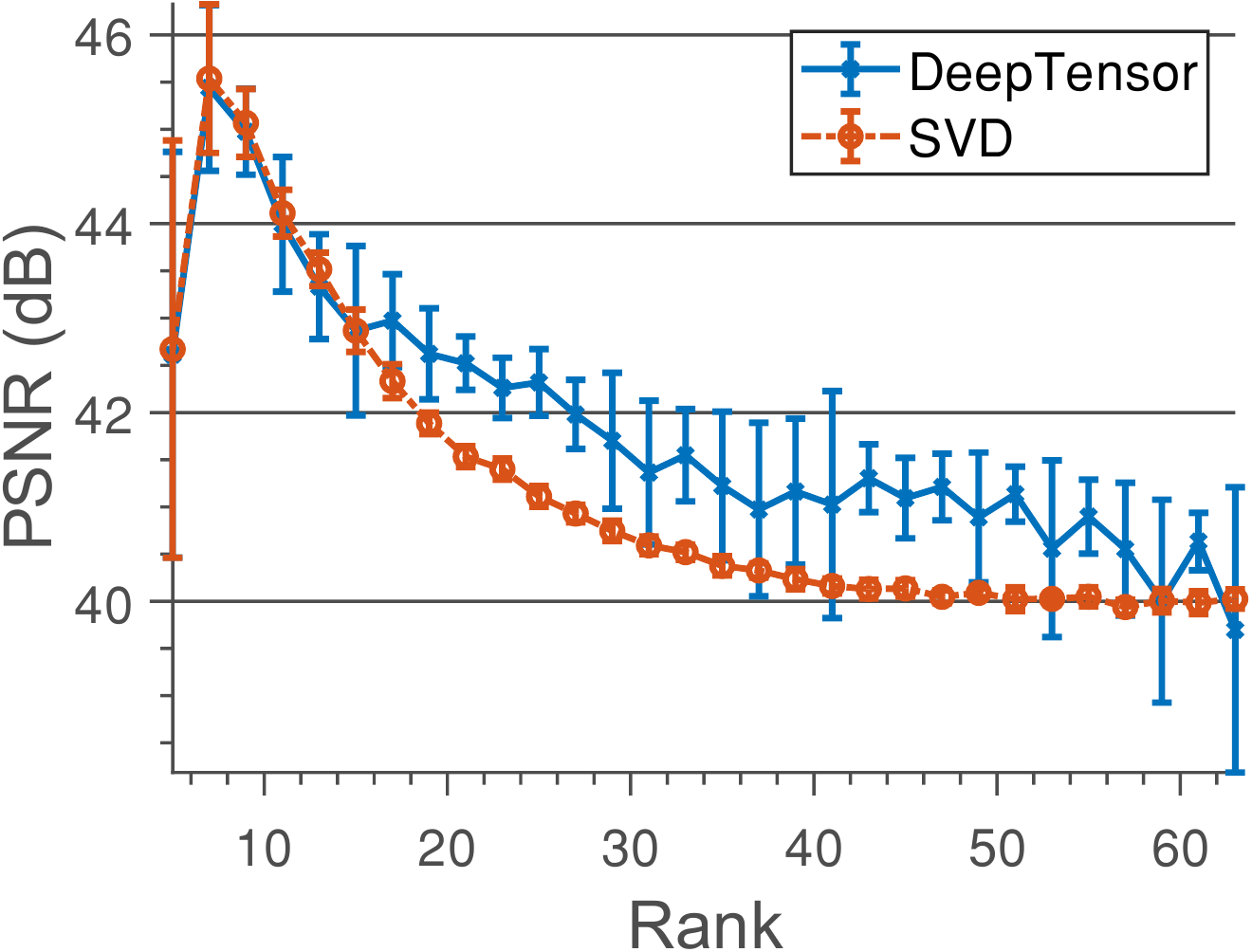}
	\end{subfigure}
	\caption{\textbf{DeepTensor is robust to a range of signal and noise distributions}. We generated low-rank matrices with various signal and noise distributions and then performed a low-rank decomposition via SVD And DeepTensor. We observe that DeepTensor has similar performance to SVD under Gaussian noise settings. When the noise is non-Gaussian, such as Poisson or Rician, DeepTensor has significantly higher accuracy across all ranks.}
	\label{fig:lr_signal_noise}
\end{figure*}

\begin{figure*}[!tt]
	\centering
	\begin{subfigure}[c]{0.31\textwidth}
		\centering
		\caption*{PCA with 32 data samples}
		\includegraphics[width=\columnwidth]{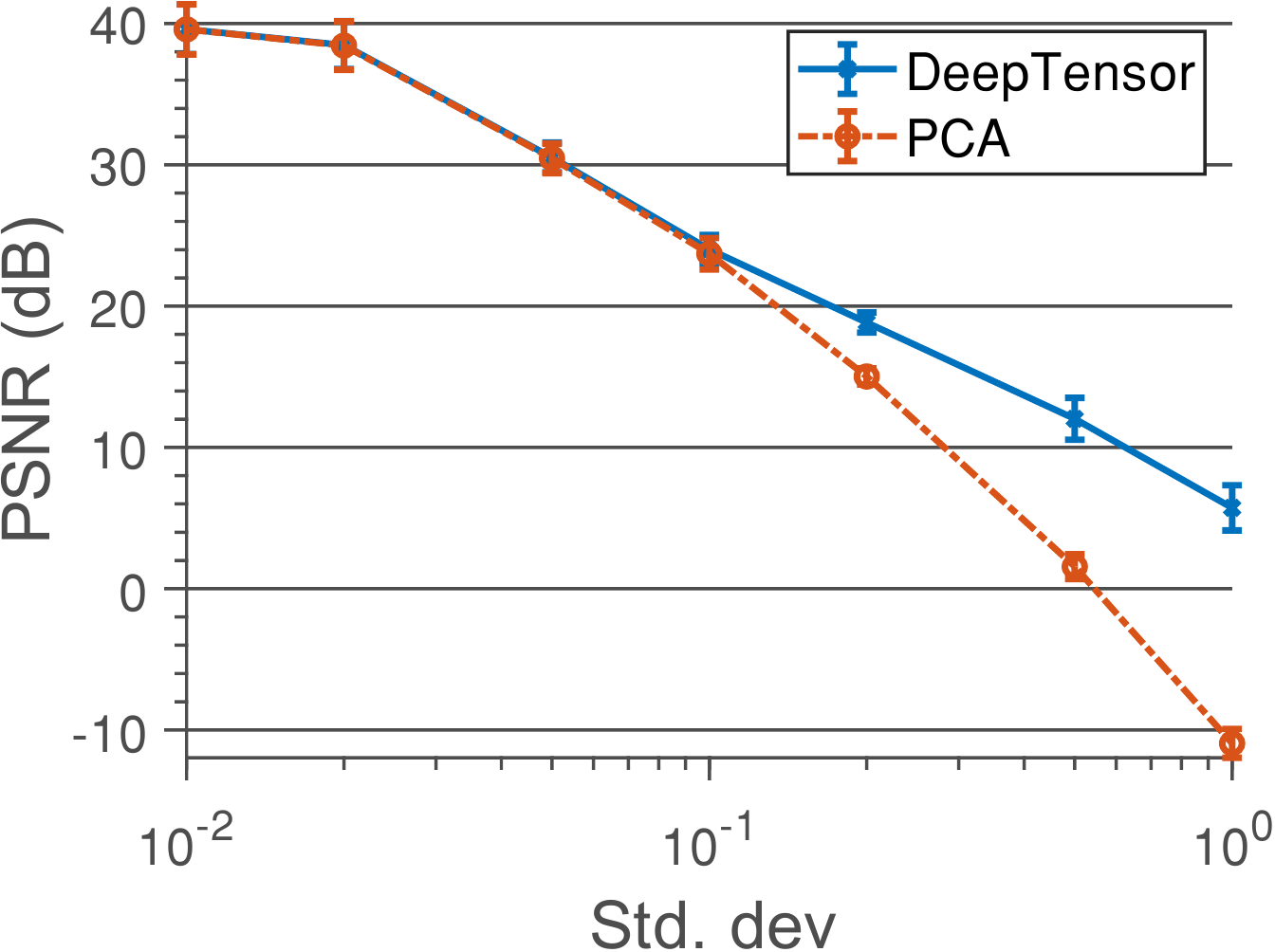}
	\end{subfigure}
	\hspace{0.5em}
	\begin{subfigure}[c]{0.31\textwidth}
		\centering
		\caption*{PCA with 1024 data samples}
		\includegraphics[width=\columnwidth]{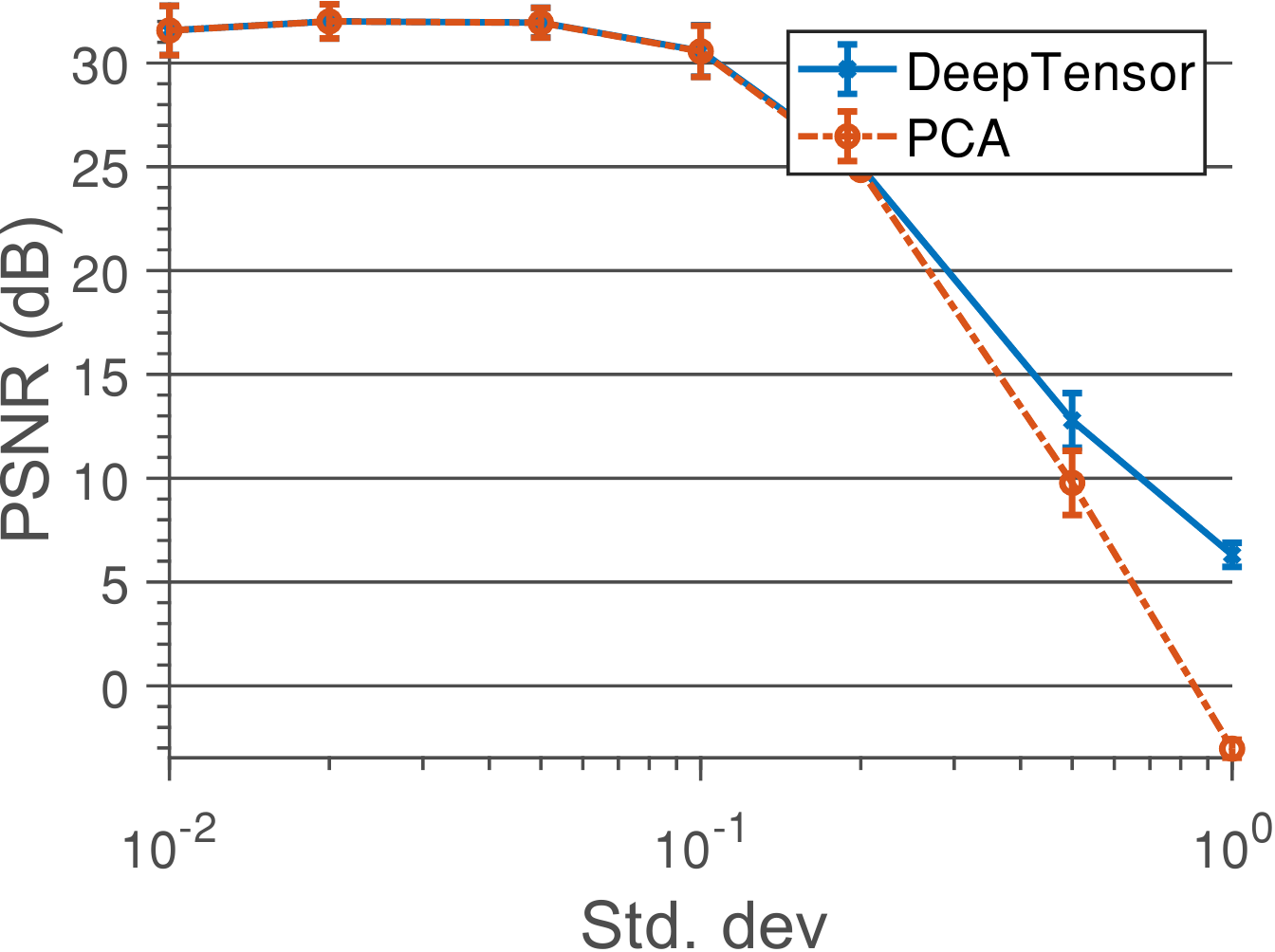}
	\end{subfigure}
	\hspace{0.5em}
	\begin{subfigure}[c]{0.31\textwidth}
		\centering
		\caption*{PSNR vs.\ sample size}
		\includegraphics[width=\columnwidth]{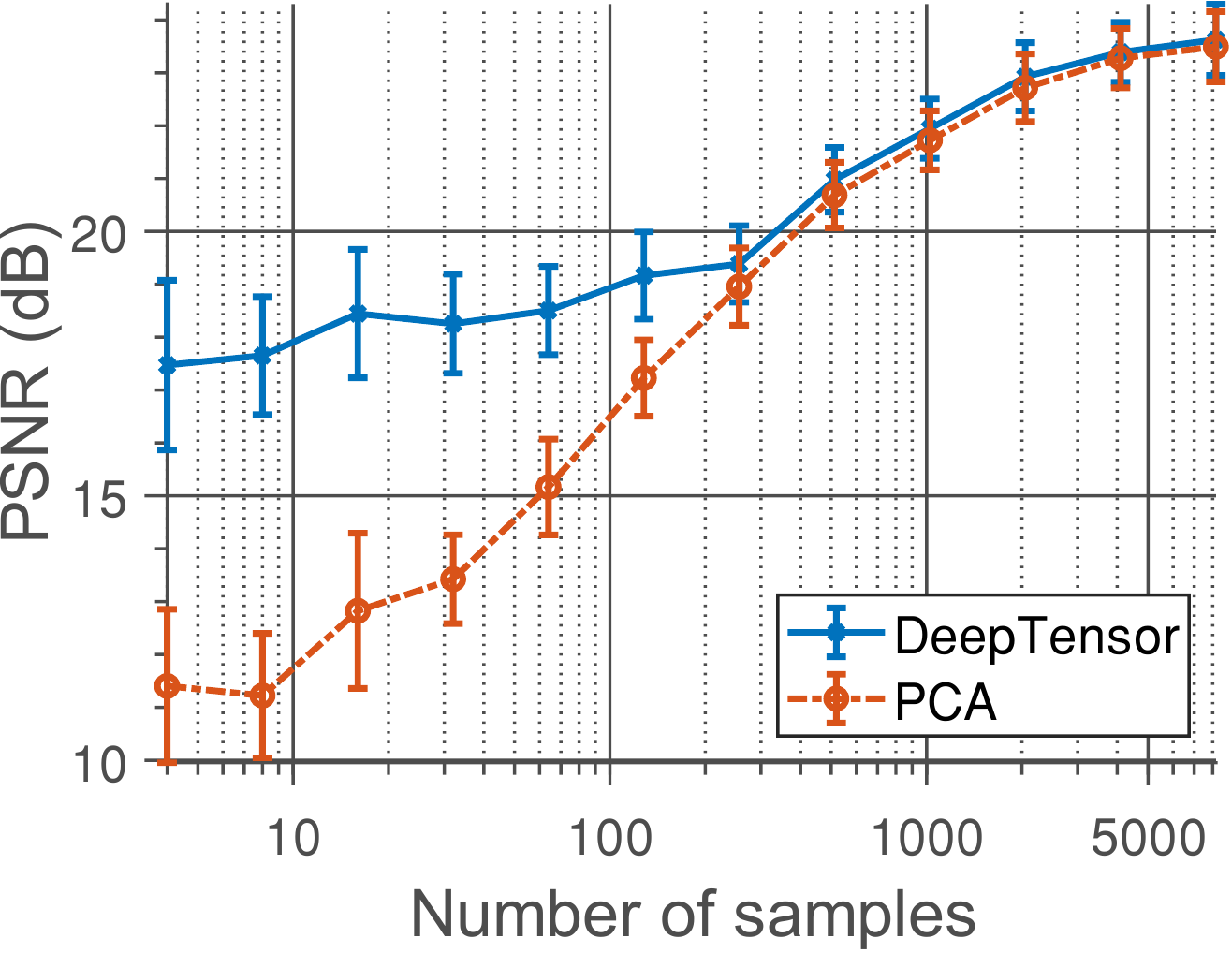}
	\end{subfigure}
	\caption{\textbf{DeepTensor is a more robust for PCA under low SNR or small sample size.}
	The SVD is the core computation of PCA. We generated data with varying number of samples drawn from a Gaussian distribution with varying standard deviation. We then computed PCA components via SVD and DeepTensor decomposition. We observe that DeepTensor is particularly advantageous when the data is corrupted by a large noise, or the number of samples are limited.}
	\label{fig:pca_ndata}
\end{figure*}

\textbf{Optimization procedure.} We optimized for the parameters of the networks that output the factor matrices using stochastic gradient descent. Specifically, we employed the ADAM optimizer~\cite{kigma14} and implemented the optimization using PyTorch framework~\cite{NEURIPS2019_9015}.
For all experiments, we optimized an $\ell_2$ loss function between the data and the product of outputs of DNs.
The primary purpose of choosing a simple loss function was to emphasize on the regularization capabilities of the DNs.
In practice, it is possible to choose a more appropriate loss function for each specific problem.

\subsection{Validation experiments}
\textbf{Low-rank completion.} We generated random $64\times64$ dimensional matrices with rank varying from 10 to 60. The left and right matrices were generated as either iid Gaussian values (with a standard deviation of 1 unit), or random, piecewise constant signals, emulating visual signals.
We then added one of the three types of noise:
\begin{itemize}
	\item iid Gaussian noise with standard deviation of 0.1 units: $Y = X + \mathcal{N}(0, 0.1)$
	\item Poisson noise with mean at each entry of the matrix lying in the range $[0, 1000]$: $Y = \mathcal{P}(1000X)$, which is common in visual signals
	\item Rician noise with standard deviation of 0.02 units: $Y = \sqrt{(X + \mathcal{N}(0, 0.02))^2 + \mathcal{N}(0, 0.02)^2}$ that is common in MRI measurements.
\end{itemize}
%

Figure \ref{fig:lr_signal_noise} shows a plot of peak signal-to-noise ratio (PSNR) as a function of rank for various signal and noise types averaged over 10 realizations.
We make two observations.
First, when the noise is Gaussian, DeepTensor has similar performance to SVD/PCA. This is expected, since SVD/PCA is known to be the optimal low-rank decomposition for white Gaussian matrices~\cite{eckart1936approximation}.
Second, for other noise settings, such as Poisson or Rician, DeepTensor has a far superior performance across all rank values.
This empirically establishes that DeepTensor is well suited for a range of non-Gaussian noise models.
\begin{table*}[!tt]
	\caption{\textbf{Effect of learning rate schedule.} The table presents the best achievable learning rate for low-rank approximation of a toy matrix. We repeated each experiment five times. SVD accuracy was $14.4$ dB. The choice affects the final achievable accuracy -- fixed scheduling with a high learning rate performs better than other choices.}
	\label{tab:lr}
	\centering
	\begin{tabular}{cllll}
		\toprule
		\backslashbox{Max learning rate}{Scheduler}& Fixed & Step & Exponential & Cosine annealing                    \\
		\midrule
		$10^{-2}$ & $17.6 \pm 0.7$ &$17.1 \pm 0.4$ &$16.6 \pm 0.4$ &$16.5 \pm 0.3$\\
		\midrule
		$10^{-3}$ & $17.5 \pm 0.6$ &$17.1 \pm 0.5$ &$16.6 \pm 0.4$ &$16.6 \pm 0.3$\\
		\midrule
		$10^{-4}$ & $17.6 \pm 0.6$ &$17.1 \pm 0.4$ &$16.6 \pm 0.4$ &$16.5 \pm 0.3$\\
		\midrule
		$10^{-5}$ & $16.5 \pm 0.3$ &$16.5 \pm 0.2$ &$16.5 \pm 0.4$ &$16.6 \pm 0.3$\\
		\bottomrule
	\end{tabular}
\end{table*}

\textbf{Principal Component Analysis.} 
%
The performance of SVD-based PCA degrades under noisy conditions or when samples are limited.
To verify how DeepTensor can benefits PCA, we generated variable number of data points with a known intrinsic matrix generated as iid Gaussian random variables with mean 0 and standard deviation of 1. 
The feature dimension was 64 and the intrinsic dimension (rank) was 10. 
We generated data via a linear combination of the columns where the weights were drawn from an iid Gaussian distribution with mean 0 and standard devation.
We then added iid Gaussian noise to the data matrix with varying levels of standard deviation.
Figure~\ref{fig:pca_ndata} plots the accuracy of estimating the PCA components for varying noise levels and number of samples averaged over 10 realizations.
We observe that with low noise or a large number of samples, SVD-based PCA and DeepTensor have similar performance.
However, with high noise or a limited number of samples, DeepTensor significantly outperforms SVD-based PCA.



\subsection{Sensitivity to learning parameters and architectures}
We now discuss how the choice of learning rate scheduling and the network architecture affect the final accuracy obtained by DeepTensor.

\textbf{Learning rate scheduling.}
The choice of learning rate and its scheduling directly affects the maximum achievable accuracy. 
To gain empirical insight into how the rate and its scheduling affect the training process, we performed a rank-16 low-rank decomposition of a $64\times64$ matrix.
%
The entries of the matrix were drawn from iid Gaussian distribution.
We then varied the learning rates from $10^{-5}$ to $10^{-2}$, and chose four learning rate schedulers, namely fixed with no change in learning rate, step scheduler where the learning rate was multiplied by $0.99$ after every 2,000 epochs, exponential scheduler with a multiplication factor of $0.9999$, and cosine annealing-based scheduling~\cite{huang2017snapshot}.

Table~\ref{tab:lr} shows results for varying learning rate and its schedule.
%
We note that a fixed scheduler results in higher PSNR across all learning rates.
Moreover, learning rates ranging from $10^{-4}$ to $10^{-2}$ are equivalent choices -- a very low learning rate of $10^{-5}$ resulted in poorer PSNR.
The two observations above imply that DeepTensor does not require complex learning rate scheduling and is robust to the learning rates.

\begin{figure}
	\centering
	\begin{subfigure}[c]{0.48\columnwidth}
		\centering
		\includegraphics[width=\columnwidth]{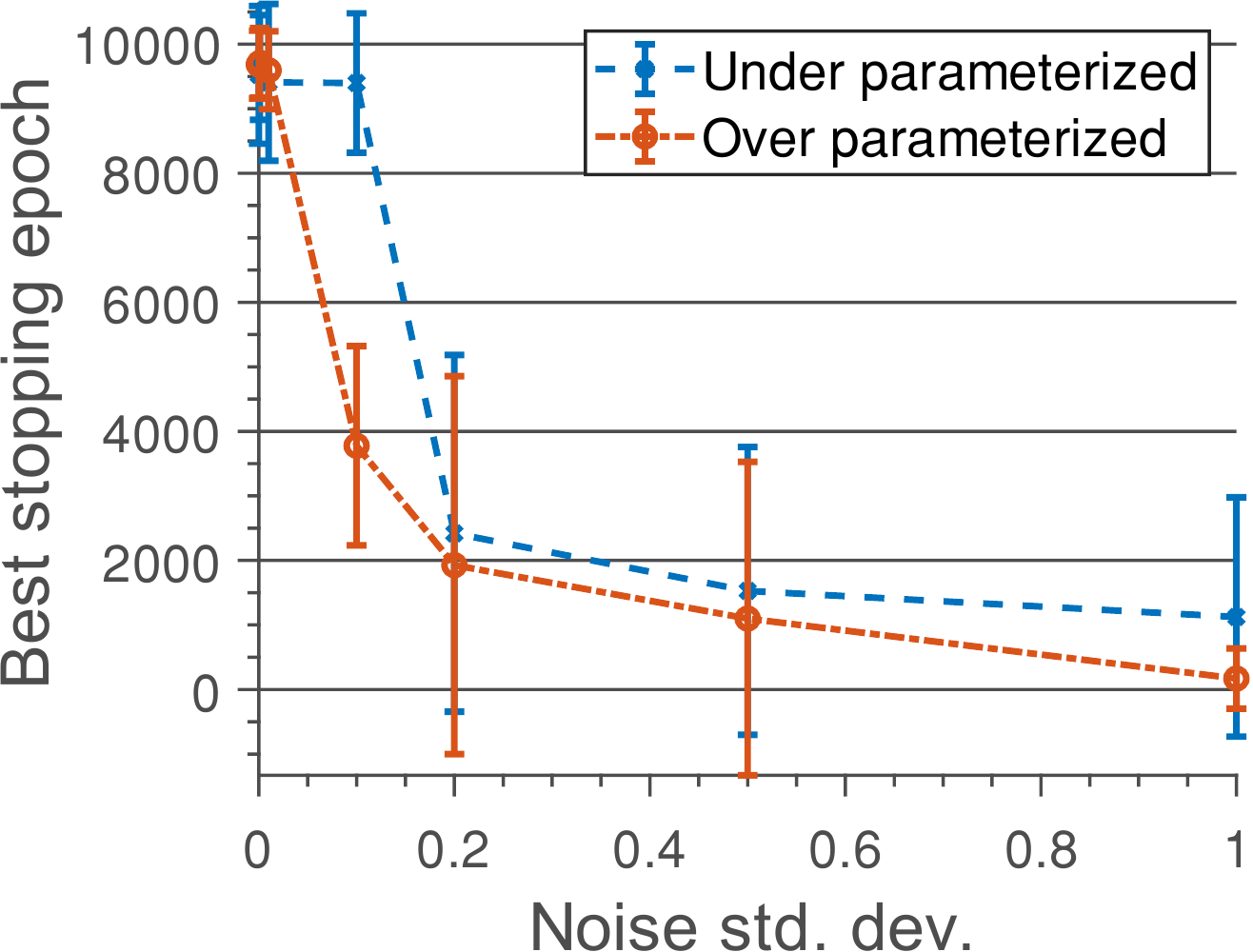}
		\caption{Best stopping epoch as a function of input noise level}
	\end{subfigure}
	\begin{subfigure}[c]{0.48\columnwidth}
		\centering
		\includegraphics[width=\columnwidth]{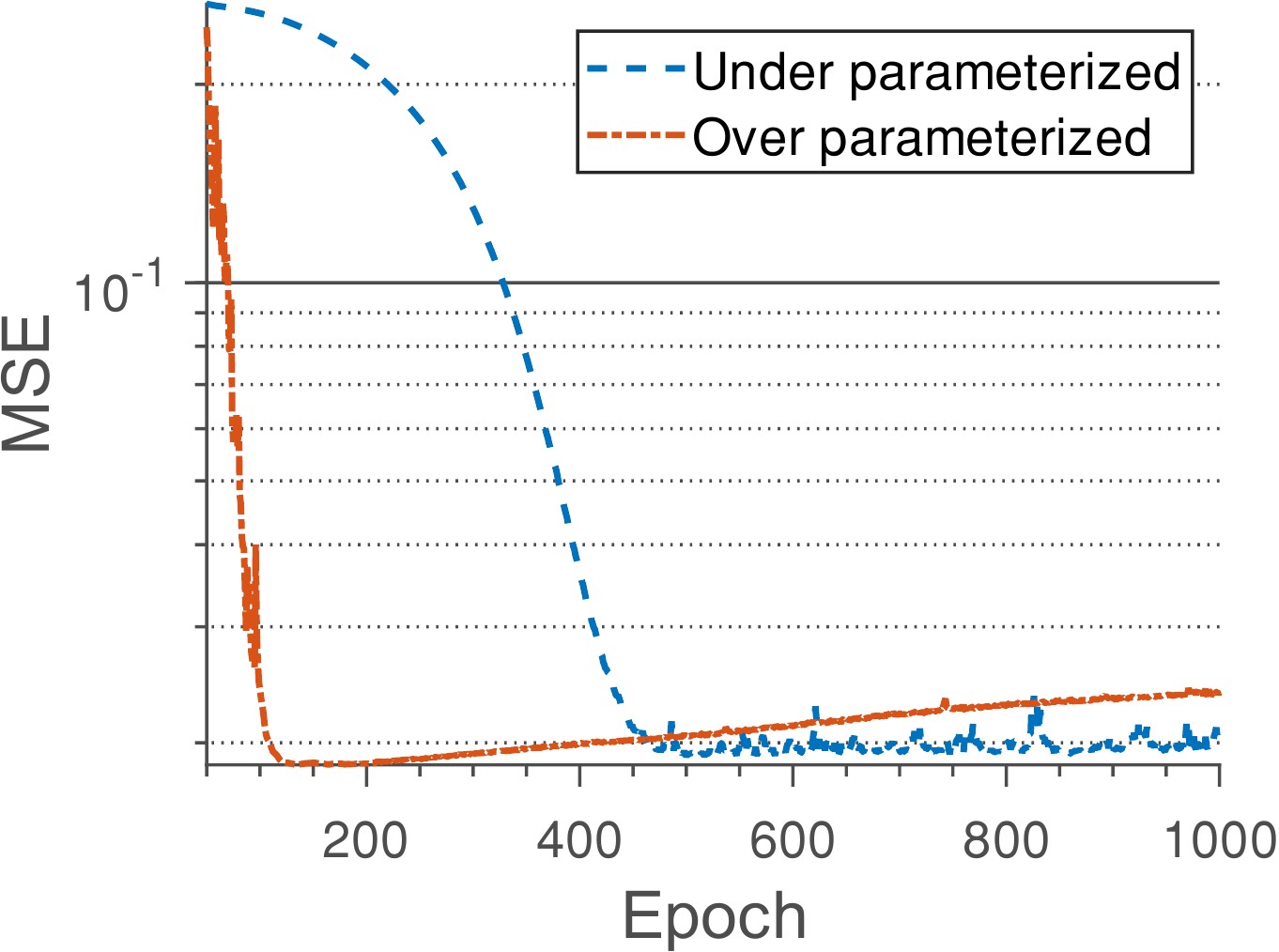}
		\caption{Training epochs for over and under-parameterized networks.}
	\end{subfigure}
	\caption{\textbf{Stopping criterion depends on input noise level and network architecture.} The optimal stopping epoch (a) reduces with increasing noise standard deviation. (b) shows the approximation error when the noise standard deviation is 0.2. The exact stopping epoch is less important for under-parameterized networks~\cite{heckel2018deep}, whereas over-parameterized networks~\cite{ulyanov2018deep} are more sensitive to the stopping epoch, but achieve lower approximation error.}
	\label{fig:dd_vs_dip}
\end{figure}

\textbf{Stopping criterion.}
%
%
%
The stopping criterion for optimal approximation accuracy is a function of input noise distribution and network architecture.
%
%
%
To demonstrate this dependence, we performed a rank-16 low-rank decomposition of $64\times64$ matrices with entries drawn from iid Gaussian distribution.
%
%
We then utilized an under-parameterized, and over-parameterized network to estimate the left and right factor matrices with varying levels of noise.
Figure~\ref{fig:dd_vs_dip} compares the results with the two types of networks.
We note in Fig.~\ref{fig:dd_vs_dip}(a) that both networks require fewer iterations with increasing noise level for least approximation error.
%
%
%
%
%
%
Fig.~\ref{fig:dd_vs_dip}(b) shows approximation error over epochs.
Over-parameterized networks achieve lower approximation error than under-parameterized networks. However,the error for over-parameterized networks increases rapidly after optimal stopping epoch (100), whereas the error increases more gently for under-parameterized networks after the optimal stopping epoch (500).
Ultimately, the stopping criteria and the network architecture depend on the exact application and prior knowledge about noise levels in the signal.
%

%
%
%

\subsection{Modeling Known Decomposition Constraints with the Last Layer Activation}
\label{sec:constraints}

Different constraints on $U$ and/or $V$ lead to different known low-rank decompositions such as nonnegativity on $U$ and $V$ leads to NMF and nonnegativity on $U$ alone leads to semi-NMF \cite{ding2008convex}, which is an ubiquitous form of clustering \cite{ding2005equivalence}. In our case, $U$ and $V$ are outputs of a DN (recall Sec.~\ref{eq:dlrp}) and are thus without range constraints. 
Hence we need to impose nonnegativity constraints on the output.
%

Nonnegativity can be achieved with various activation functions including ReLU, softplus, and element-wise absolute value, and the exact choice affects the achievable accuracy.
To gain an empirical insight into the the effect of the activation function, we considered NMF on MNIST~\cite{lecun1998gradient} and CIFAR~\cite{krizhevsky2009learning} datasets.
In both cases, we used 2048 images for training. We added a rician noise to the images of the form $\bfn = 0.3(0.3\bfz_1 + \bfz_2^2)$ where $\bfz_1$ and $\bfz_2$ are iid Gaussian random variables with zero mean and unit variance. 
The input PSNR was evaluated to be $4.8$ dB.
We then performed NMF on the resultant images with various techniques. Comparisons are tabulated in Table~\ref{tab:activations}.
%
%
We employed a combination of $\ell_2$ loss for data fidelity, and $\ell_1$ penalty for the factor matrices output from the DNs.
%
%
DeepTensor, particularly when combined with the ReLU activation function achieves higher approximation accuracy compared to baseline NMF algorithm~\cite{cichocki2009fast}.

\begin{table}[!tt]
	\caption{\textbf{DeepTensor is a robust alternative to NMF.} We performed NMF on MNIST and CIFAR10 datasets by enforcing nonnegativity on $U, V$ through different activation functions. The table below shows the average PSNR over 10 runs. DeepTensor with ReLU as the final activation function outperformed standard NMF~\cite{cichocki2009fast}, underscoring the efficacy for matrix factorization with positivity constraints.}
	\label{tab:activations}
	\centering
	\begin{tabular}{llll}
		& \multicolumn{3}{c}{\textbf{MNIST}}                   \\
		\cmidrule(r){2-4}
		\textbf{Act. func.}  & softplus     & abs. value & ReLU \\
		\midrule
		DeepTensor & 7.4 $\pm$0.03& 7.3$\pm$0.01& $\mathbf{7.6\pm0.15}$ \\
		\midrule
		NMF~\cite{cichocki2009fast} & \multicolumn{3}{c}{7.5}\\
		\bottomrule
	\end{tabular}
	\begin{tabular}{llll}
		& \multicolumn{3}{c}{\textbf{CIFAR10}}                   \\
		\cmidrule(r){2-4} 
		\textbf{Act. func.}  & softplus     & abs. value & ReLU \\
		\midrule
		DeepTensor & 8.4 $\pm$0.08 & 8.3$\pm$0.06 & $\mathbf{8.8\pm0.17}$\\
		\midrule
		NMF~\cite{cichocki2009fast} &\multicolumn{3}{c}{8.2}\\
		\bottomrule
	\end{tabular}
\end{table}








\section{Applications}\label{section:experiments}
We now showcase the breadth of DeepTensor's applicability in several real-world applications. 
%

\textbf{Training details.}  Unless otherwise specified, we used overparameterized networks with skip connections (similar to DIP~\cite{ulyanov2018deep}), and trained with a learning rate of $10^{-3}$. We implemented our training process in Python using the Pytorch framework~\cite{NEURIPS2019_9015}.
%
%
All our experiments were run on a Linux machine equipped with a hexa-core Intel Core i7-6850K CPU, 128GB RAM, and three NVIDIA GeForce GTX 1080 Ti.
The network architectures were a modified version of the one used by \citet{ulyanov2018deep}. 
Specifically, we changed the number of output channels to be equal to the rank for all factor matrices.
For tensor decomposition with 1D fibers, we changed the 2D convolutional architecture in \cite{ulyanov2018deep} to a 1D architecture.
We optimized for both inputs and network parameters, which provided faster convergence; however, there was no other significant difference if we did not optimize for the inputs.
%

\subsection{Linear inverse problems in computer vision}
Low-rank model finds use in numerous applications in computer vision including sensing of light transport matrices~\cite{o2010optical}, hyperspectral imaging~\cite{saragadam2018krism}, video compressive sensing~\cite{hitomi2011video}, magnetic resonance imaging (MRI), and positron emission tomography (PET).
Most inverse problems involve collection of limited data samples and/or highly noisy samples. In both cases, we expect the DeepTensor to be very effective.
We showcase three specific examples here.

\textbf{Hyperspectral image denoising.} Low-rank models offer a concise representation of hyperspectral images (HSI) and are used in compression, sensing~\cite{saragadam2018krism}, and dimensionality reduction.
HSIs involve imaging the scene across several hundreds of spectral bands, resulting in high photon noise (highly non-Gaussian).
Typically, a HSI of dimension $N_x \times N_y \times N_\lambda$ is converted to a matrix of dimension $N_x N_y \times N_\lambda$ which is then approximated using a low-rank model.
We denoised a $348\times327\times260$ HSI from \citet{arad_and_ben_shahar_2016_ECCV} by simulating Poisson noise equivalent to a maximum of 100 photons per spatio-spectral voxel, and a readout noise of 2 photons -- settings corresponding to a dull overcast outdoor scene.
We then performed a rank-20 decomposition with SVD and DeepTensor.
We also compared DeepTensor against the BM3D denoising algorithm~\cite{dabov2007image}.
We ran the DeepTensor optimization process for a total of 5000 iterations.
Figure~\ref{fig:lr_signal_noise} shows the results with both techniques.
DeepTensor outperforms SVD by 6dB, and BM3D by 3dB and produces visually pleasing results.

\begin{figure*}[!tt]
	\centering
	\includegraphics[width=\textwidth]{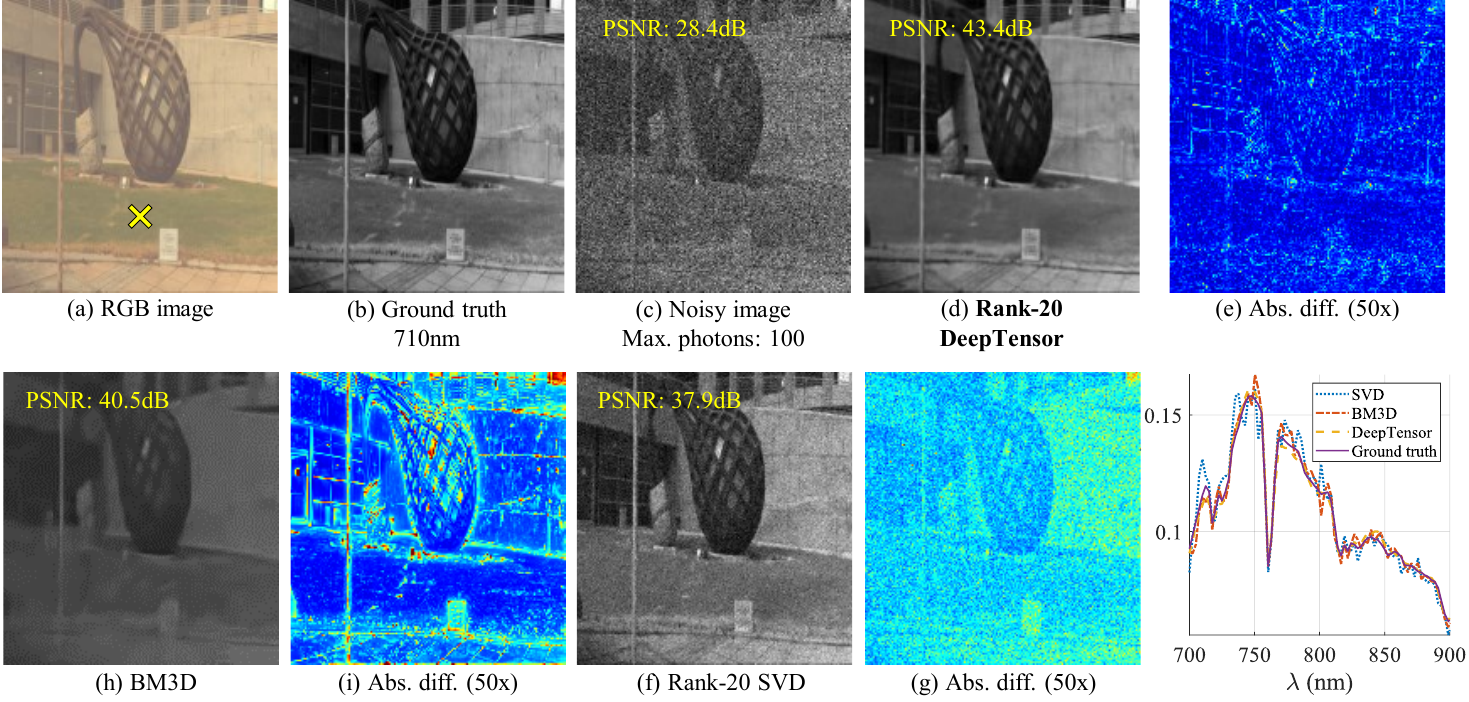}
	\caption{\textbf{Hyperspectral denoising.} Low-rank models are often used for representing hyperspectral images. We simulated noisy hyperspectral images by adding Poisson noise with a maximum mean of 100 counts and then denoised using a rank-20 model. DeepTensor has approximately 6dB increase in performance over SVD and 3dB over BM3D~\cite{dabov2007image}.
	}
	\label{fig:hyperspectral}
\end{figure*}

\begin{figure}
    \centering
    \includegraphics[width=\columnwidth]{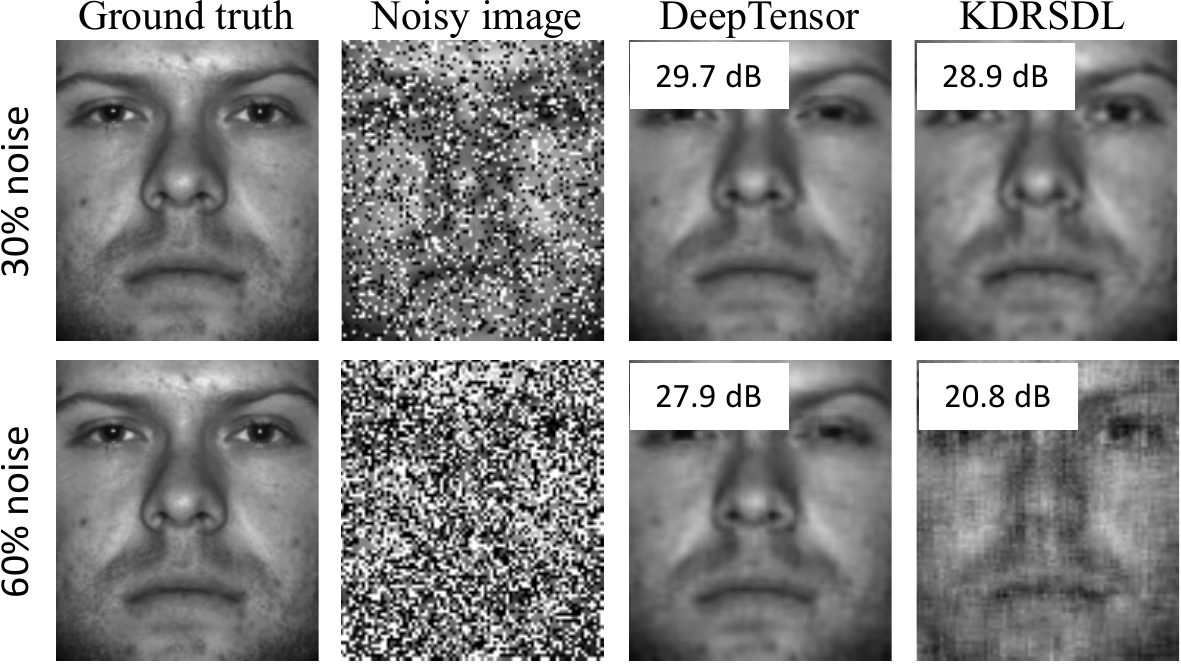}
    \caption{\textbf{Tensor denoising.} DeepTensor enables denoising via low-rank tensor representation that is often better than state-of-the-art denoising techniques~\cite{bahri2018robust}.}
    \label{fig:kdrsdl}
\end{figure}

\begin{figure*}[!tt]
	\centering
	\includegraphics[width=\textwidth]{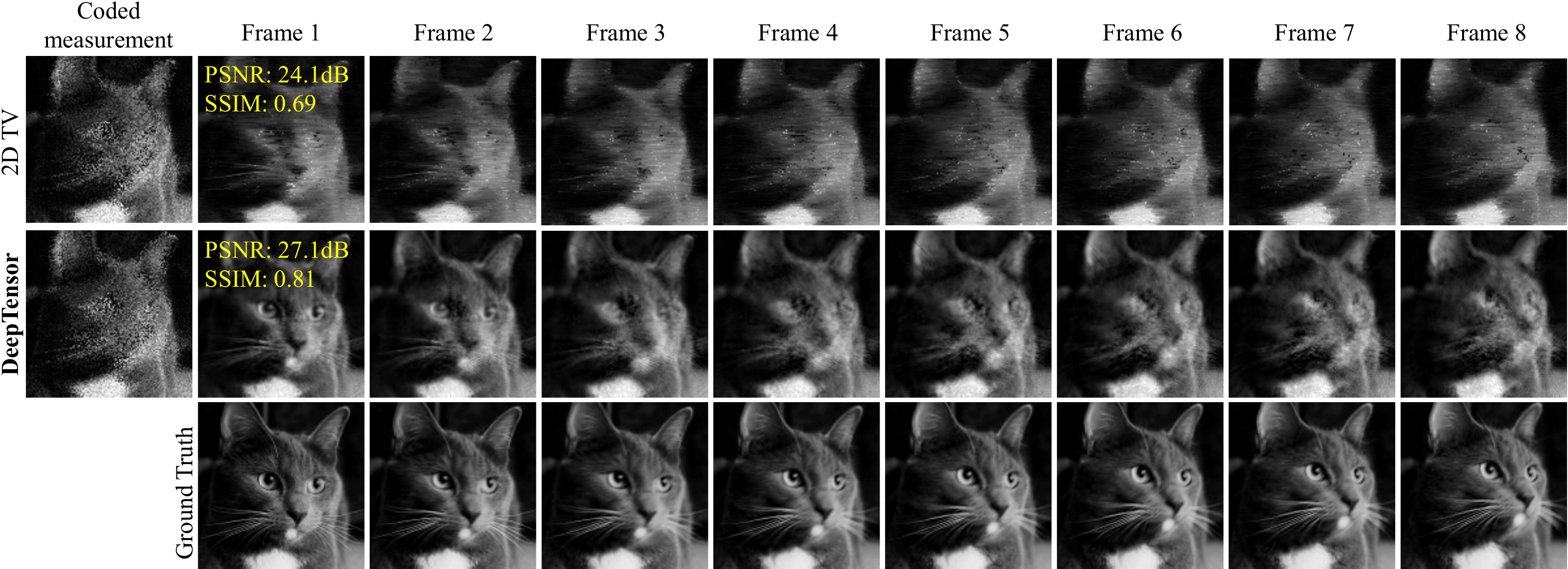}
	\caption{\textbf{Video compressive sensing.} DeepTensor can be used for solving linear inverse problems with compressed measurements. We simulated video CS following the setup by \cite{hitomi2011video} to combine 32 frames into a single coded image frame. We then recovered with DeepTensor at full rank, and with 2D TV. DeepTensor produced significantly higher quality of results.}
	\label{fig:video}
\end{figure*}

\begin{figure*}[!tt]
	\centering
	\includegraphics[width=\textwidth]{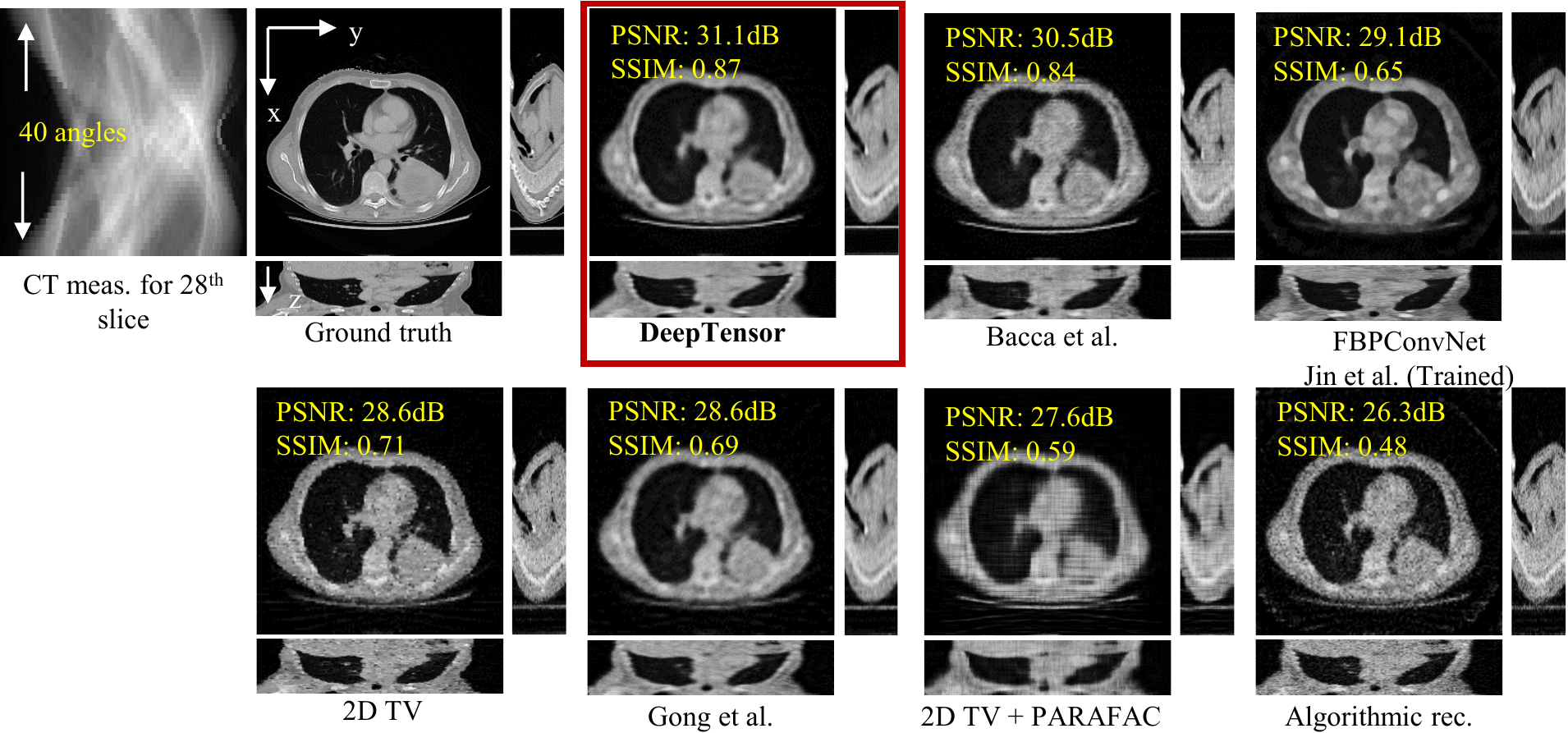}
	\caption{\textbf{Reconstruction from 3D CT measurements.} We utilized DeepTensor to recover 3D PET volume from limited, noisy CT images. We expressed the 3D volume as a rank-1000 PARAFAC tensor and then computed CT images per each slice along 40 angles. DeepTensor resulted in higher SSIM compared to other approaches. }
	\label{fig:pet}
\end{figure*}

\textbf{Tensor denoising.} Low dimensional tensor decomposition is a promising approach to tensor denoising, especially with gross outliers such as salt and pepper noise.
We performed a 3-way PARAFAC decomposition of a subset of faces from the Yale face database (B)~\cite{belhumeur1997eigenfaces} consisting of 160, $192\times168$-dimensional images.
We then added $30\%$, and $60\%$ salt and pepper noise for a fair comparison against the Kronecker Decomposition-based approach (KDRSDL) which is the state-of-the-art in tensor decomposition~\cite{bahri2018robust}.
We found that an $\ell_1$ penalty worked better than $\ell_2$ for DeepTensor, as the noise was spatially sparse.
In both cases, we picked a rank of 2000 for PARAFAC decomposition.
Results are shown in Fig.~\ref{fig:kdrsdl}.
Evidently, DeepTensor outperforms KDRSDL, especially in extremely noisy settings.

\textbf{Video compressive sensing.} DeepTensor can also be used for full-rank decomposition, which is apt for signals such as videos.
To test this, we used our framework on recovery of video frames from spatially multiplexed images.
We relied on the setup from \citet{hitomi2011video} where each pixel was sampled at an arbitrary time frame across multiple frames. We combined 8 frames into one single coded image and then used DeepTensor to solve the linear inverse problem.
For comparison, we recovered video by solving the linear inverse problem with a 2D TV over spatial images as a regularizer.
We trained DeepTensor for a total of 10,000 epochs.
Figure~\ref{fig:video} shows coded image as well as the 8 recovered images for an example video.
DeepTensor not only has higher accuracy than TV, but the recovered images look visibly similar to ground truth.
%

\textbf{3D reconstruction from CT images.} 
DeepTensor is well suited for tasks such as recovery of 3D volume from CT scans.
To demonstrate the advantages of using a tensor representation, we rely on PARAFAC decomposition.
We approximated a $256\times256\times56$ PET CT scan from \citet{clark2013cancer} with a rank-1000 PARAFAC tensor. 
Note that although the rank is much larger than any single dimension, the effective number of parameters are only $15\%$ of the number of elements in the 3D volume.
We added Poisson and readout noise of maximum of 100 photons per voxel and 2 photons respectively.
We then simulated 40 tomographic measurements for each slice.
2D TV results were obtained by a TV penalty on each slice along z-direction. 
\citet{bacca2021compressive} results were obtained by representing input as a rank-1000 PARAFAC tensor and using an untrained 2D network which output 56 channels.
2D TV + PARAFAC results were obtained with self-supervised learning by representing the volume via rank-1000 PARAFAC decomposition, and a TV penalty on each slice along z-axis.
Algorithmic reconstruction results were obtained by solving the linear inverse problem without any other priors.
FBPConvNet~\cite{jin2017deep} results were obtained by using the output of algorithmic reconstruction and then denoising with a pre-trained model.
Finally, the results by \cite{gong2018pet} were obtained with a self-supervised learning by using a 3D convolutional neural network. While \cite{gong2018pet} utilized an MRI image as input for the network, we used uniform random noise.
Results are visualized in Fig.~\ref{fig:pet}.
DeepTensor achieves better reconstruction accuracy than other approaches in both PSNR and SSIM.
We note that pre-trained approaches such as FBPConvNet~\citet{jin2017deep} can outperform DeepTensor if trained with the appropriate data.
%


\subsection{Classification via low-dimensional projection}
A robust low-rank approximation also affects downstream tasks such as classification.
To test this, we worked with an Eigenfaces~\cite{turk1991eigenfaces} example for facial classification.
We took 840 images across 28 subjects from the Weizmann dataset~\cite{weizmann}. We used $25\%$ of the data for training an 84-dimensional subspace via PCA, Independent Component Analysis (ICA)~\cite{comon1994independent}, and running DeepTensor on the sample covariance matrix.
To emulate noisy conditions, we added poisson noise with variable amounts of noise.
We then converted the images to the 84-dimensional space and trained linear and kernel support vector machine with radial basis function.
We cross-validated to choose penalty that maximized classification accuracy for each individual classifier.
Finally, we evaluated the resultant linear subspace along with SVM on the test data to compute average accuracy.
Figure~\ref{fig:pca} visualizes the learned basis vectors with the three approaches, and the average classification accuracy as a function of input data PSNR.
We notice that the basis learned from DeepTensor is smoother and better representative of the underlying data.
In contrast, PCA and ICA overfit to the noise in data, resulting in a reduction in classification accuracy.






\subsection{Time-frequency decomposition}
As discussed in section 3.4, DeepTensor can be combined with additional constraints such as nonnegativity on $U, V$. 
We validated the advantages of DeepTensor by performing an NMF on speech data obtained from the GOOGLECOMMANDS dataset~\cite{warden2018speech}.
We first computed spectrograms on all speech data with a window size of $1024$ samples and a hop size of $32$, resulting in a $512\times512$ dimensional time-frequency image.
We then added noise as described in section 3.4, resulting in an input SNR of $4.6$ dB. 
The approximation accuracies with DeepTensor, and a baseline NMF~\cite{ding2008convex} NMF algorithm in Table~\ref{tab:activations_TFR}.
DeepTensor performs significantly better than the baseline.
This ability of DeepTensor to perform better in high noise settings is of particular significance when speech is recorded in noisy environments.
%
%

\subsection{Choice of rank}
\begin{figure}
    \centering
    \begin{subfigure}[t]{0.48\columnwidth}
		\centering
		\includegraphics[width=\columnwidth]{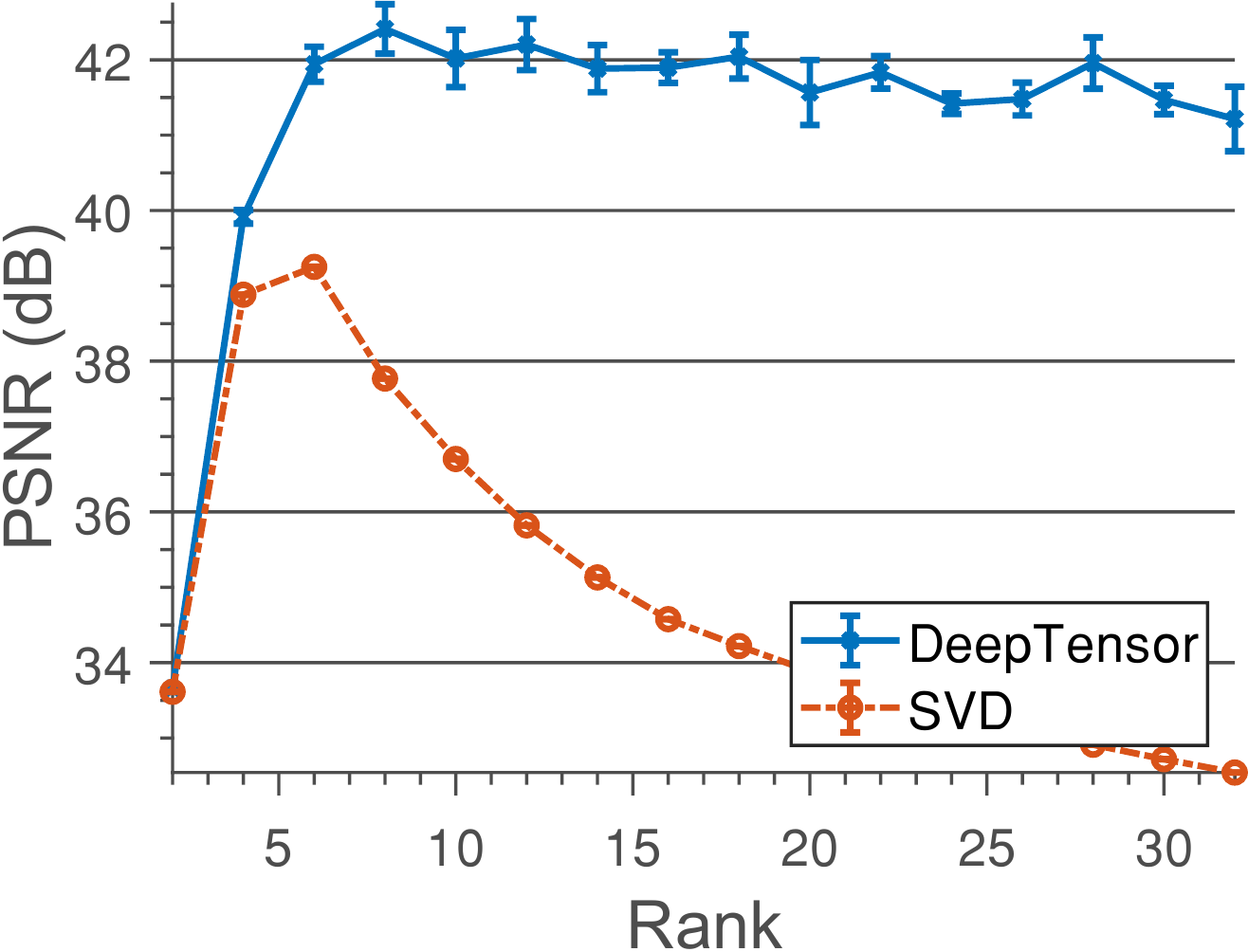}
		\caption{Matrix decomposition}
	\end{subfigure}
	\begin{subfigure}[t]{0.48\columnwidth}
		\centering
		\includegraphics[width=\columnwidth]{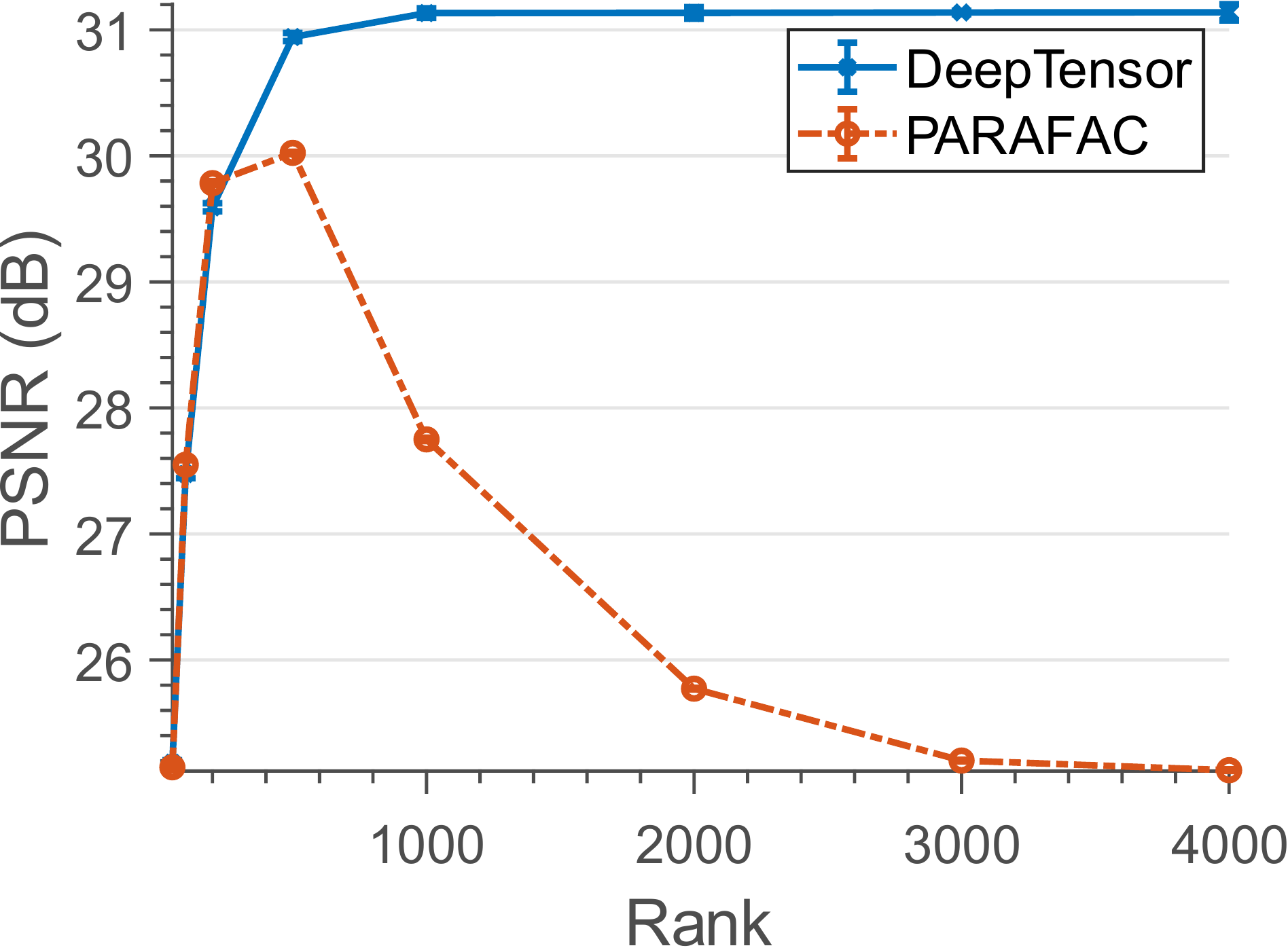}
		\caption{PARAFAC tensor decomposition}
	\end{subfigure}
    \caption{\textbf{Effect of choice of rank.} We truncated a hyperspectral image to rank-20 and a tensor to rank-1000 (PARAFAC) and added shot noise.  We then swept rank of decomposition. We observe that DeepTensor is less sensitive to rank compared to SVD and PARAFAC.}
    \label{fig:psnr_vs_rank}
\end{figure}

As with all matrix and tensor decomposition approaches, the rank of the decomposition is an important parameter for DeepTensor.
However, we now demonstrate that DeepTensor is less sensitive to the choice of rank than the SVD in two different settings.
%
%
In the first experiment, we truncated a hyperspectral image~\cite{deepcassi2017} to rank-20 and swept the rank from 1 to 30.
In the second experiment, we truncated a subset of the Yale-B dataset~\cite{belhumeur1997eigenfaces} using rank-1000 PARAFAC decomposition and then swept the decomposition rank from 10 to 4000.
In both cases, we added Poisson ($\lambda_\text{max} = 100$) and Gaussian noise ($\sigma = 2$) equivalent to a 25 dB measurement PSNR.
Figure~\ref{fig:psnr_vs_rank} shows plots of PSNR as a function of rank for both the cases.
When the rank is under-estimated, DeepTensor and SVD achievely similarly low accuracy.
However, when the rank is \emph{over-estimated}, the achievable accuracy with SVD reduces, while the accuracy with DeepTensor stays approximately the same.
The exact choice of rank is less important for DeepTensor compared to the SVD, which is significantly advantageous when the appropriate rank to use is unknown.
\begin{table*}[!tt]
	\caption{\small \textbf{NMF on speech data.} Average and standard deviation (over $10$ runs) of the PSNR (dB) for GOOGLECOMMANDS~\cite{warden2018speech} in the NMF (nonnegativity of $U, V$) and semi-NMF (nonnegativity of $U$ alone) low-rank decomposition setting for different activation functions enforcing nonnegativity. DeepTensor with ReLU performs better than standard NMF~\cite{cichocki2009fast}.}
	\label{tab:activations_TFR}
	\centering
	\begin{tabular}{lllllll}
		\toprule
		& \multicolumn{3}{c}{\textbf{GOOGLECOMMANDS NMF}} &\multicolumn{3}{c}{\textbf{GOOGLECOMMANDS SEMI-NMF}}                   \\
		\cmidrule(r){2-4} \cmidrule{5 - 7}
		\textbf{Act. func.}  & softplus     & abs. value & ReLU & softplus & abs. value & ReLU \\
		\midrule
		DeepTensor & 9.0 $\pm$0.03& 8.7$\pm$0.01 & 8.7$\pm$0.03& 8.7$\pm$0.09& 8.8$\pm$0.02
		&8.8$\pm$0.02\\
		\midrule
		baseline~( \cite{cichocki2009fast,ding2008convex}) & \multicolumn{3}{c}{4.6}&\multicolumn{3}{c}{4.7}\\
		\bottomrule
	\end{tabular}
\end{table*}

\section{Conclusions}\label{section:conclusion}
\textbf{Discussion.} 
%
We have demonstrated that self-supervised learning is effective for solving low-rank tensor and matrix decomposition.
Across the board, we see that DeepTensor is a superior option compared to SVD/PCA when the input SNR is low, the matrix/tensor values are non-Gaussian distributed, or the inverse problem is ill-conditioned such as in the case of PCA with limited samples, or linear inverse problems.
Moreover, our separable approximation approach results in faster approximation of 3D tensors compared to DNs with 3D convolutional filters.
%

\textbf{Future directions.}
Our experiments relied mostly on 1D or 2D convolutional filters, whose inductive biases are better suited for signals such as time series and images.
However, the DNs can be chosen specific to the task at hand, such as fully connected or recurrent networks.
%
%
DeepTensor can also be used in non-linear representations, such as local low-rank, or even non-local low-rank. Such settings increase the range of problems that can be tackled effectively and are exciting future directions.

\textbf{Limitations.} The current bottleneck of DeepTensor, as opposed to techniques such as SVD, lies in the training computational complexity. 
While a single pass through a DN is much faster than the training of SVD or NMF, the need to repeatedly iterate between the forward passes and back propagation inherent to gradient based learning slows DeepTensor training. %
This opens up interesting research directions aimed at discovering simpler networks that are faster to train yet maintain high performance and developing specialized training algorithms that leverage the low-rank decomposition structure.

\section{Acknowledgments}\label{section:ack}
This work was supported by NSF grants CCF-1911094, IIS-1838177, and IIS-1730574; ONR grants N00014-18-12571, N00014-20-1-2534, and MURI N00014-20-1-2787; AFOSR grant FA9550-22-1-0060; and a Vannevar Bush Faculty Fellowship, ONR grant N00014-18-1-2047

{\small
	\bibliographystyle{IEEEtranN}
	\bibliography{refs}
}

%

\end{document}